\documentclass[letterpaper,twocolumn,showpacs,floatfix,groupaddress,longbibliography]{revtex4-1}
\usepackage[latin1]{inputenc}
\usepackage{bm}
\usepackage{multirow}
\usepackage{amssymb}
\usepackage{amsbsy}
\usepackage{amsmath,  amsthm, amsfonts, mathrsfs}
\usepackage{graphicx}
\usepackage{epsfig}
\usepackage{placeins}
\usepackage{float}
\usepackage{trfsigns}
\usepackage[usenames,dvipsnames]{color}
\usepackage[colorlinks,linkcolor=blue,citecolor=blue,urlcolor=blue,breaklinks=true]{hyperref}

\makeatletter

\begin{document} 

 \title{Compelling Bounds on Equilibration Times - the Issue with Fermi's Golden Rule}
 
 \author{Robin Heveling}
 \email{rheveling@uos.de}
 \affiliation{Department of Physics, University of Osnabr\"uck, D-49069 Osnabr\"uck, Germany}

 \author{Lars Knipschild}
 \email{lknipschild@uos.de}
 \affiliation{Department of Physics, University of Osnabr\"uck, D-49069 Osnabr\"uck, Germany}

 \author{Jochen Gemmer}
 \email{jgemmer@uos.de}
 \affiliation{Department of Physics, University of Osnabr\"uck, D-49069 Osnabr\"uck, Germany}

\begin{abstract}
Putting a general, physically relevant upper bound on equilibration times in closed quantum systems is a recently much pursued endeavor. 
In PRX, 7, 031027 (2017) Garc\'{\i}a-Pintos et al. suggest such a bound. We point out that the general assumptions which allow for an actual estimation of this bound are violated
%this bound  is practically inconclusive 
in cases in which Fermi's Golden Rule 
and related open quantum system
theories apply.  To probe the range of applicability of Fermi's Golden Rule for  systems of the type addressed in the above work, we numerically solve the
corresponding Schr\"odinger equation for some finite spin systems comprising up to 25 spins.  These calculations shed light on the breakdown of standard quantum master
equations 
in the ``superweak'' coupling limit, which occurs for finite sized  baths.

\end{abstract}

 %---------------------------------------------------------------------------------------

\maketitle

\section{Introduction}
The last decades have seen a major progess in the field of equilibration in closed quantum systems \cite{gogolin16}. Concepts like typicality  \cite{cantyp,reimann,lloyd} and the 
eigenstate thermalization hypothesis \cite{schrecknicki,deutsch}
have been brought forth. Furthermore, it has been established that for an initial state $\rho$ populating  many energy levels, 
expecation values $\langle A(t) \rangle$ will generically be very close to their temporal averages for most times within the interval to which the average refers
(``equilibration on average'').
While the conditions for this statement to be true are rather mild concerning the observable $A$ and the Hamiltonian $H$ \cite{reimann2,linden,4}, 
the respective time interval may be very large. For specific 
observables it may, e.g., scale with the dimension of the relevant Hilbert space \cite{goldstein,malabarba}. Moreover, concrete examples are known in which the corresponding equilibration times for physically relevant 
observables scale as 
$T_\text{eq}\propto N^\alpha$, $\alpha \geq 1/2$, where $N$ is the size of the system. This result has been found for systems featuring long range \cite{kastner}
as well as short range interactions \cite{5}, albeit for a somewhat different definitions of equilibration times. In fact, already for 
mesoscopic many-body systems with standard interaction strengths, the required equilibration  interval may be on the order of the age of the universe \cite{kastner}. 
Thus, although the above statements in some sense establish
equilibration under moderate conditions in the very long run, it is unclear whether or not this equilibration will ever occur in a 
physically relevant period of time.
Hence, the question of an upper bound on this relaxation timescale has recently been much discussed. Since it is always possible to find mathematically 
well defined, permissible initial states that fully exhaust the above, unsatisfactorily large time interval, most contributions focus on additional, 
physically plausible conditions. These conditions, which are intended to capture the actual, physical state of affairs, 
may be imposed on the initial state, the observable, the structure of the system, or combinations thereof \cite{cramer,diez,vinayak,torres}. In the present paper we primarily discuss results from 
Ref. \cite{garcia-pintos17}. The latter rest on assumptions on all of the above.\\
\noindent
This paper is organized as follows: In Sect. \ref{garferm} we briefly present a main result from Ref. \cite{garcia-pintos17}. Furthermore, we elaborate on the lack 
of predicitive power of this result in  cases in which   Fermi's Golden Rule applies. In Sect. \ref{spinexp} we explain some  models, each of which comprises a single
spin in a magnetic field interacting with a (finite) bath, consisting of spins itself. We initialize the system in a standard 
system-bath product state and numerically solve the Schr\"odinger equation, monitoring the system's spin component parallel to the magnetic field. These data unveil the regime of 
validity of Fermi's Golden Rule with respect to the crucial system parameters. 
Sect. \ref{scaltheo} discusses the scaling of the critical interaction strength at which open system predictions start to become unreliable.
In Sect. \ref{discussion} the implications of the numerical findings from Sect.  \ref{spinexp} on the 
assumptions and statements from  Ref. \cite{garcia-pintos17} are named and explained. Eventually, we sum up and conclude in Sect. \ref{sumcon}.
\vspace*{-15px}
\section{Garc\'{\i}a-Pintos bound and Fermi's Golden Rule}
\label{garferm}
\vspace*{-5px}
To begin with, we state a main result of Ref. \cite{garcia-pintos17} (hereafter called the Garc\'{\i}a-Pintos bound (GPB)) in a comprehensive form. 
The GPB addresses an equilibration time $T_\text{eq}$. To further specify $T_\text{eq}$ we introduce some notation. Let $\rho$ be the initial state of the system. Let 
furthermore $A(t)$ denote an observable $A$ in the Heisenberg picture and $\langle A(t) \rangle := $ Tr$[A(t)\rho]$ its time dependent expectation value. Due to 
the closed system dynamics being unitary (and the system being finite),   $\langle A(t) \rangle$ has a well defined ``infinite time average'' $\overline{ \langle A \rangle}:= \langle A \rangle_\text{eq} $,   which is routinely 
considered as the equilibrium value of 
$A$ in case the observable $A$ equilibrates at all \cite{4}. Consider now a deviation $D(t)$ of the actual expectation value from its equilibrium, i.e., 
$D(t):= (\langle A(t) \rangle - \langle A \rangle_\text{eq})^2/ 4 ||A||^2$, with $||A||$ being the largest absolute eigenvalue of $A$. Consider furthermore an average of  $D(t)$ over the time interval $[0,T]$
denoted by $\overline{D\vphantom{l}}_{T}$. The condition that defines $T_\text{eq}$ is that $\overline{D\vphantom{l}}_{T} \ll 1$ must hold for $T\gg T_\text{eq}$ (for non-equilibrating systems such a
 $T_\text{eq}$ may not exist \cite{4}). The GPB is an 
explicit expression
for such a $T_\text{eq}$ (see Eq. (\ref{gpb})), based on $\rho, A(0)$ and $H$, where $H$ is the Hamiltonian of the system. 
As the GPB involves somewhat refined  functions of the above three operators, we need to specify these before stating the GPB explicitly. A central role takes a kind of probability 
distribution $p_{jk}$ which is defined as
\begin{eqnarray}
    \label{palfa}
    p_{jk} &\propto& |\rho_{jk}A_{kj}| \;\,\text{for} \,\; E_j-E_k \neq 0\,, \\
    p_{jk} &=& 0  \;\,\text{for} \,\; E_j-E_k = 0, \quad      \sum_{j,k}  p_{jk} = 1\,, \nonumber
\end{eqnarray}
where $E_j, E_k$ are energy eigenvalues corresponding to energy eigenstates $|j\rangle , |k\rangle$. Furthermore, matrix elements are 
abbreviated as $\rho_{jk}:=\langle j| \rho| k \rangle, A_{jk}:=\langle j |A(0) |k \rangle$. While the GPB is not limited to this case, we focus here on $p_{jk}$ which allow for a description in terms of a probability 
density function $w(G)$. All examples we present below conform with such a description and it is plausible that this applies to many generic many-body scenarios.
Prior to defining $w(G)$, we define $w(G,\epsilon)$ as
\begin{equation}
    \label{wege}
    %w(G,\epsilon) := \frac{1}{\epsilon} \sum_ {G' :   G-\epsilon/2 \leq   G' \leq   G+\epsilon/2 }   p_{G'}. 
    w(G,\epsilon) := \frac{1}{\epsilon} \sum_ {j,k} \Theta \! \left(  \frac{\epsilon}{2} - |E_j-E_k -G| \right)      p_{jk}\,,
\end{equation}
where $\Theta$ is the Heaviside function. This is the standard construction of a histogram in which the $p_{jk}$ are sorted according to their
respective energy differences $E_j-E_k$. It is now assumed that there exists a range of (small but not too small) $\epsilon$ such that $ w(G,\epsilon)$
is essentially independent of  variations of $\epsilon$ within this range. The   $ w(G,\epsilon)$ from this ``independence regime'' are simply
abbreviated as 
$w(G)$.
Let the standard deviation of $w(G)$ be denoted 
by $\sigma_G$. Let furthermore $w_\text{max}$ denote the  maximum of $w(G)$. The quantities $a$ and $Q$ that eventually enter the GPB are now defined as 
\begin{equation}
\label{defaq}
a:= w_\text{max}\sigma_G\,, \quad \quad Q:= \sum_{i,j: E_i \neq E_j} \dfrac{|\rho_{ij} A_{ji}|}{||A||}\,.
\end{equation}
We are now set to state the GPB:
\begin{equation}
\label{gpb}
T_{\text{eq}} = \dfrac{\pi a ||A||^{1/2} Q^{5/2}}{\sqrt{|\vphantom{\dot{h}}\text{Tr}([[\rho,H],H]A)|}} =  
\dfrac{\pi a ||A||^{1/2} Q^{5/2}}{\sqrt{|\vphantom{\dot{h}}\frac{\text{d}^2}{\text{d}t^2}\langle A(t) \rangle\big{|}_{t=0}|}} \,.
\end{equation}
Obviously, the GPB links $T_{\text{eq}}$ to the initial ``curvature'' of the observable dynamics $\partial_t^2\langle A(t)\rangle|_{t=0}$ (which is practically accessible, cf. Fig. \ref{fig7}). An actual, concrete bound on the equilibration time by means of $T_{\text{eq}}$, however, only arises from Eq. (\ref{gpb}) if the numerator can be shown to be in an adequate sense small or at least bounded. This is a pivotal feature on which the ``predictive power'' of the GPB hinges. The crucial quantities in the numerator are $a$ and $Q$. As it is practically impossible to calculate $a$ from its definition for many-body quantum systems, Garc\'{\i}a-Pintos et al. instead offer an assumption. \\
They argue that $a \sim 1$ may be expected for  $w(G)$ that are ``unimodal''. Unimodal means 
that 
$w(G)$ essentially consists of one central elevation like a Gaussian or a box distribution, etc. Indeed, $a$ is invariant with respect to a rescaling as 
$w(G) \rightarrow s w(sG)$, as it would result from rescaling the Hamiltonian as $H  \rightarrow sH$ (here $s$ is some real, positive number).
Garc\'{\i}a-Pintos et al. also offer various upper bounds on $Q$ for different situations.\\
\noindent
In the remainder of this section, we explain in which sense the conclusiveness of the GPB is in conflict with Fermi's Golden Rule (FGR). Let us stress that this conflict does not concern the validity or correctness of Eq. (\ref{gpb}) as such, the latter is undisputed. It only concerns the assumptions on $a$ and $Q$, which are required to find an actual value or estimate for $T_{\text{eq}}$. (Note that there is some evidence (cf. Sect. \ref{discussion}) that specifically the assumption on $a$  is violated, rather than the assumption on $Q$). Consider an Hamiltonian consisting of an unperturbed part $H_0$
and a perturbation $H_{\text{int}}$.
\begin{equation}
\label{ham}
H = H_0 + \lambda H_{\text{int}}
\end{equation}
Consider furthermore an observable $A$, which is conserved under $H_0$, i.e. $[A, H_0]=0$. If $ H_0 $ has a sufficiently wide and dense spectrum and $\lambda$ is small, 
FGR may apply under well investigated conditions \cite{hove,bartsch,kupsch}.  The applicability of the FGR approach yields, in the simplest case, 
a monoexponential decay, i.e.
\begin{equation}
\label{decay}
\langle A(t) \rangle = (\langle A(0) \rangle -\langle
A\rangle_{\text{eq}})\e^{-t/\tau_\text{rel}} +\langle
A\rangle_{\text{eq}}\,,
\end{equation}
where $\tau_\text{rel} := r \lambda^{-2}$ and $r$ is a real, positive number depending on $ H_0$ and $H_{\text{int}}$.
More refined approaches, such as the Weisskopf-Wigner theory or open
quantum system approaches, also
 arrive at such exponential decay dynamics \cite{scully,breuer}. In the
relevant case  $\partial_t  \langle A(t) \rangle|_{t=0}=0$, obviously
Eq. (\ref{decay}) cannot apply at
 $t=0$. In this case Eq. (\ref{decay}) is meant to apply
after a short ``Zeno time'' $\tau_\text{zeno}$ that
is often very short compared to the relaxation time $\tau_\text{rel}$
\cite{kupsch}. (Note, however,  that the denominator of Eq. (\ref{gpb})
addresses 
a time below the Zeno time, if the latter is nonzero). We now aim at
finding the principal dependence of quantities in  Eq. (\ref{gpb}) on the
interaction strength $\lambda$. While  the definition of $T_{\text{eq}}$
as given at the beginning of the present Sect. does not fix the relation of
$\tau_\text{rel}$ and $T_{\text{eq}}$ rigorously, for exponential decays
it appears plausible to require at least
\begin{equation}
\label{times}
T_{\text{eq}} \geq \tau_\text{rel}\,.
\end{equation}
For the denominator of   Eq. (\ref{gpb}) we find with Eq. (\ref{ham})
\begin{equation}
\label{scalden}
\sqrt{\bigg{|}\frac{\text{d}^2}{\text{d}t^2}\langle A(t) \rangle\big{|}_{t=0}\bigg{|}}
=\sqrt{|c_1\lambda + c_2\lambda^2  |}\,,
\end{equation}
where \mbox{$c_1 = \text{Tr}([ H_\text{int}, A][\rho,H_0]),       c_2 =
\text{Tr}([ H_\text{int}, A] [\rho,H_\text{int}]) $.}
Plugging Eqs. (\ref{decay}, \ref{times}, \ref{scalden})
into Eq. (\ref{gpb}) yields
\begin{equation}
\label{scalnum}
\pi a ||A||^{1/2} Q^{5/2} \geq r \frac{\sqrt{|c_1\lambda + c_2\lambda^2 
|}}{\lambda^2}
\end{equation}
for the numerator of Eq. (\ref{gpb}).
Obviously, the numerator of Eq. (\ref{gpb}) diverges in the limit of weak interactions, i.e. $\lambda \rightarrow 0$. The latter holds even if $c_1 =0$.  
This contradicts the central assumption behind the 
GPB as outlined below Eq. (\ref{gpb}). Hence, the validity of FGR in the weak coupling limit and a conclusive applicability of the GPB are mutually exclusive. 
This is the first main result
of the present paper. Although the practical success
of FGR is beyond any doubt, the theoretical applicability of FGR rests on various assumptions on the system in question, 
so does the applicability of standard open system methods. 
In order to learn about the 
applicability of either the GPB or FGR from considering examples, we analyze some spin systems  in the following Sect. \ref{spinexp} by numerically solving the respective 
Schr\"odinger equations. This analysis is comparable to numerical investigations performed in Ref. \cite{garcia-pintos17}. However, other than Garc\'{\i}a-Pintos et al. we 
analyze the weak coupling limit and consider system sizes that are too large to allow for  numerically exact diagonalization of the respective Hamiltonians.
\section{Numerical Spin-based Experiments Probing Equilibration Times}
\label{spinexp}
\begin{figure}[H]
	\centering
  \includegraphics[width=0.35\textwidth]{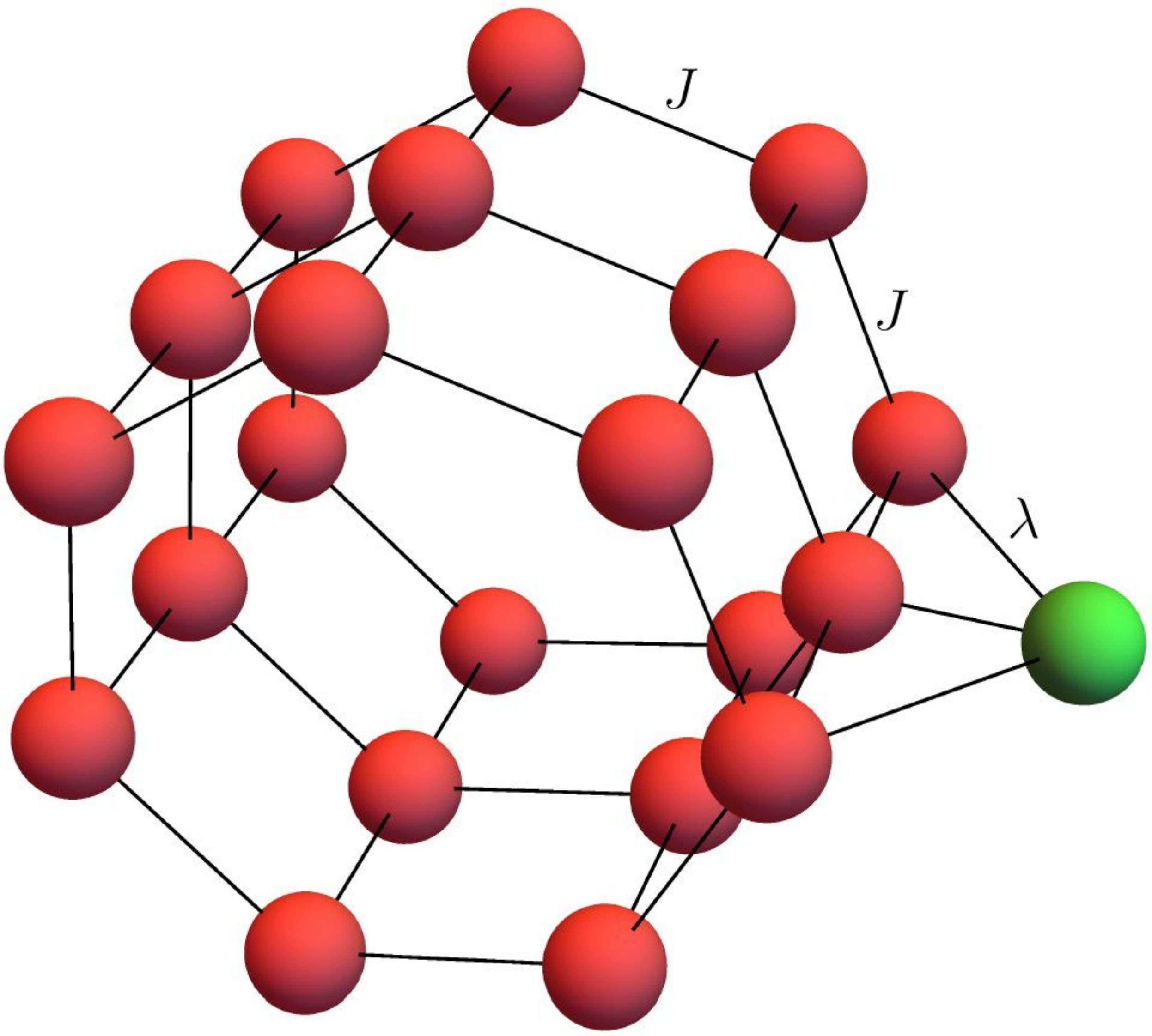}
	\caption{Single system spin (green) and spin-bath (red) interact with strength $\lambda$. Solid black lines indicate isotropic Heisenberg interactions.}
	\label{fig1}
\end{figure}

While we analyze  a number of concretely specified models below, it is important to note that these models just represent some generic instances of the 
system-bath scenarios which are routinely considered in open quantum system theory. The (non-integrable) baths share some properties with standard solid state systems, like 
periodicity and locality (in this respect they differ from the otherwise comparable models addressed in Refs. \cite{1,2,3}). Other than that, the 
details of our modeling are not peculiar at all. We varied details of the bath Hamiltonians in piecemeal fashion and found all below results unaltered (cf. App. \ref{gene}). Our archetypal model is an isotropic spin-$1/2$ Heisenberg system consisting of a single system spin coupled to a bath. The bath is rectangularly shaped with $3 \times L$ spins and features periodic boundary conditions in the longitudinal direction resulting in a wheel-like structure (cf. Fig. \ref{fig1}). Thus, the total number of spins is given by $N=3L+1$. The single system spin is subject to an external magnetic field in the $z$-direction and interacts with three neighboring bath spins in the transverse direction. This model is non-integrable in the sense of the Bethe-Ansatz. 
The bath Hamiltonian reads
\begin{equation*}
H_{\text{bath}} = J \sum_{r=1}^3 \sum_{i=1}^{L} \big{(}S^x_{i,r}S^x_{i+1,r}+S^y_{i,r}S^y_{i+1,r}+S^z_{i,r}S^z_{i+1,r}\big{)}
\end{equation*}
\vspace*{-15px}
\begin{equation}
\begin{split}
&+ J \sum_{i=1}^L  \big{(}S^x_{i,1}S^x_{i,2}+S^y_{i,1}S^y_{i,2}+S^z_{i,1}S^z_{i,2}\big{)}\\&+ J \sum_{i=1}^L  \big{(}S^x_{i,2}S^x_{i,3}+S^y_{i,2}S^y_{i,3}+S^z_{i,2}S^z_{i,3}\big{)}\,,
\end{split}
\end{equation}
where $S^{x,y,z}_{i,r}$ are spin-$1/2$ operators at site $(i,r)$ and ${L +1 \equiv 1}$. The exchange coupling constant $J$ as well as $\hbar$ are set to unity.
\noindent
The Hamiltonian of the system is given by
\begin{equation}
H_{\text{sys}} = B S^z_{\text{sys}}\,,
\end{equation}
where $S^{x,y,z}_{\text{sys}}$ denote the spin-$1/2$ operators of the additional system spin and $B=0.5$. The interaction between bath and system is described by the Hamiltonian
\begin{equation}
\begin{split}
H_{\text{int}} =\bigg{[}\big{(}&S^x_{1,1}+ S^x_{1,2}+ S^x_{1,3} \big{)}S^x_{\text{sys}} \\+ \big{(}&S^y_{1,1}+ S^y_{1,2}+ S^y_{1,3} \big{)}S^y_{\text{sys}} \\+ \big{(}&S^z_{1,1}+ S^z_{1,2}+ S^z_{1,3} \big{)}S^z_{\text{sys}}\bigg{]}
\end{split}
\end{equation}
and contributes with a factor $\lambda$ to the total Hamiltonian
\begin{equation}
H = H_{\text{sys}} + H_{\text{bath}} + \lambda H_{\text{int}}\,.
\end{equation}
The considered initial states are product states of a system state $\pi_{\uparrow}$ and a bath state $\pi_{E,\delta}$. This corresponds to a 
situation where system and bath are   initially uncorrelated and then brought into contact via $H_{\text{int}}$ at $t=0$. 
The system state is a projector onto the $S^z_{\text{sys}}$-eigenstate corresponding to spin-up.
The bath state $\pi_{E,\delta}$ is a projector onto a (small) energy window of width $\delta$ centered around a mean energy $E$. 
\begin{equation}
\label{ini}
\rho = \dfrac{\pi_{\uparrow} \otimes \pi_{E,\delta}}{\text{Tr} \{ \vphantom{\tilde{\big{(}}} \pi_{\uparrow} \otimes \pi_{E,\delta}\}} 
\end{equation}
Concretely, we fix the width of the energy window \mbox{$\delta =0.1$}, which is very small compared to the scale of the full energy spectrum of the bath. 
Given the size of the systems it comprises nevertheless a very large number of energy eigenstates.
To keep track of finite size effects we increment the baths circumference $L$ in steps of size one, thus adding three spins to the bath in each step.
\begin{figure}[H]
	\centering
  \includegraphics[width=0.48\textwidth]{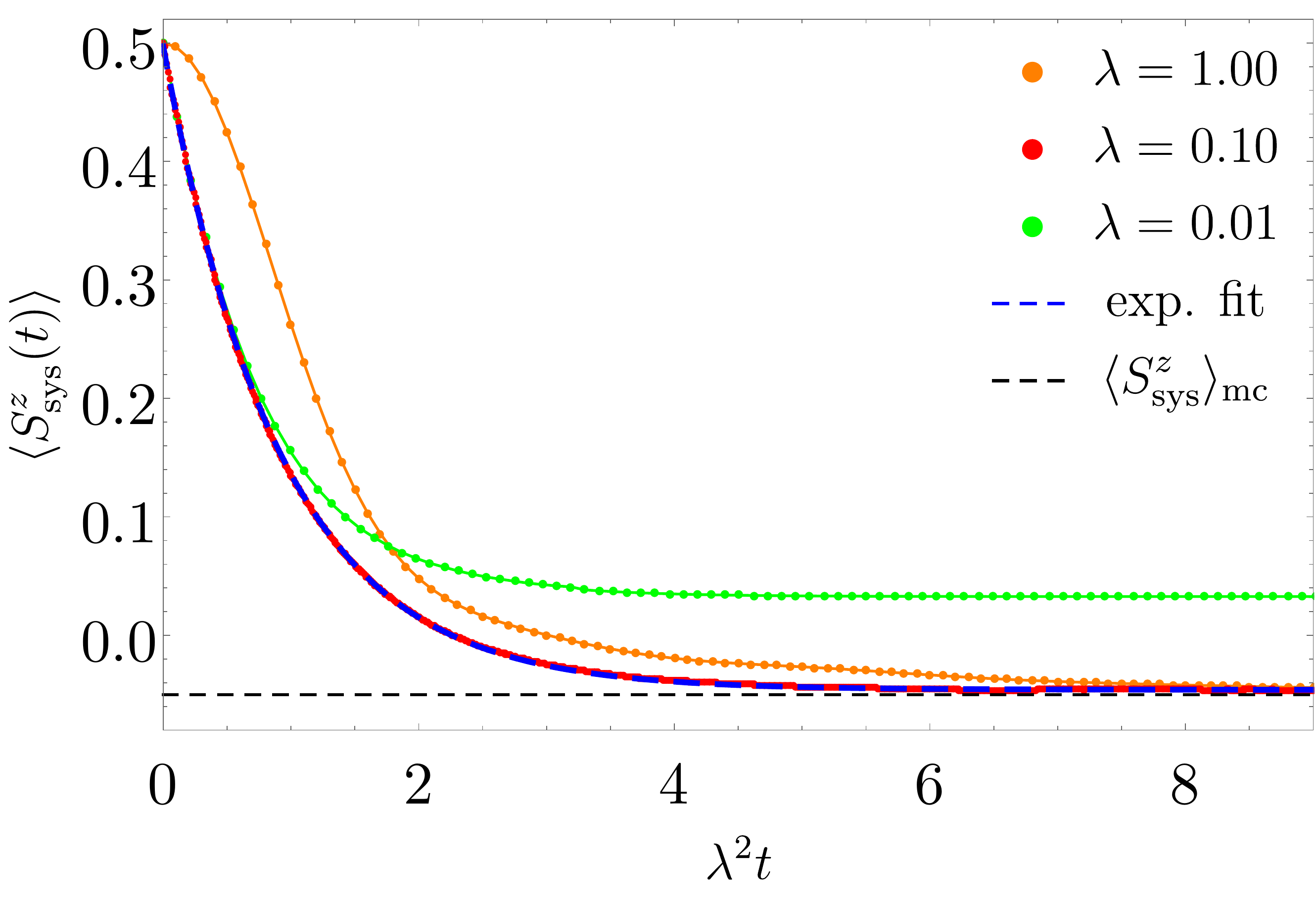}
	\caption{Decay of the magnetization for $N=25$. For strong coupling (e.g. $\lambda = 1.0$) the magnetization decays quickly and nonexponentially
	to the thermal expectation value. Note that the time axis is scaled with $\lambda^2$. For weak coupling (e.g. $\lambda = 0.1$) 
	the magnetization decays exponentially towards the thermal equilibrium value. For very weak coupling (e.g. $\lambda=0.01$) the magnetization
	gets stuck at a non-thermal longtime average value.}
	\label{fig2}
\end{figure}
\noindent
An inverse temperature $\beta$ is defined as $\partial_E \log \Omega (E)$, where $\Omega(E)$ is the density of states of the bath at energy $E$. This ``microcanonical''
definition of temperature is also employed in Ref. \cite{garcia-pintos17}. For comparability of different bath sizes
we aim at keeping  $\beta$ fixed while 
incrementing the bath size. As $H_{\text{bath}}$ is local, the bath energy is expected to 
scale linearly with 
the bath size. Hence, we choose a scaling of the initial bath energy as $E \approx - 0.15 (N-1)$, which corresponds to choosing $\beta \approx 0.4$. Given these specifications of the Hamiltonian and the initial state, we 
numerically solve the corresponding Schr\"odinger equation and monitor the expectation value of the $z$-component of the magnetization of the system-spin, i.e.
$\langle S^z_{\text{sys}}(t) \rangle$. Some results are displayed in Fig. \ref{fig2} for a schematic overview. In accord with open quantum system theory, these results suggest to 
distinguish three cases. \\i. {\em non-Markovian regime:} For strong coupling (e.g. $\lambda = 1.0$) the magnetization quickly decays to the
 equilibrium value, i.e. $\langle S^z_{\text{sys}}\rangle_\text{mc} $, in a nonexponential 
way. The description of these dynamics requires the incorporation of memory effects in some way. While this is a very active field in open quantum theory, we do not 
investigate this regime any further in the present paper.\\
ii.  {\em Markovian regime:} For weak coupling (e.g. $\lambda = 0.1$) there is a monoexponential decay to the thermal equilibrium. 
This exponential decay is in full accord with  FGR. A large number of systems ranging from quantum optics to condensed matter fall into this regime \cite{breuer, weiss}. 
It also largely coincides with the field of quantum semi-groups and the Lindblad approach.\\

\begin{figure}[H]
	\centering
  \includegraphics[width=0.48\textwidth]{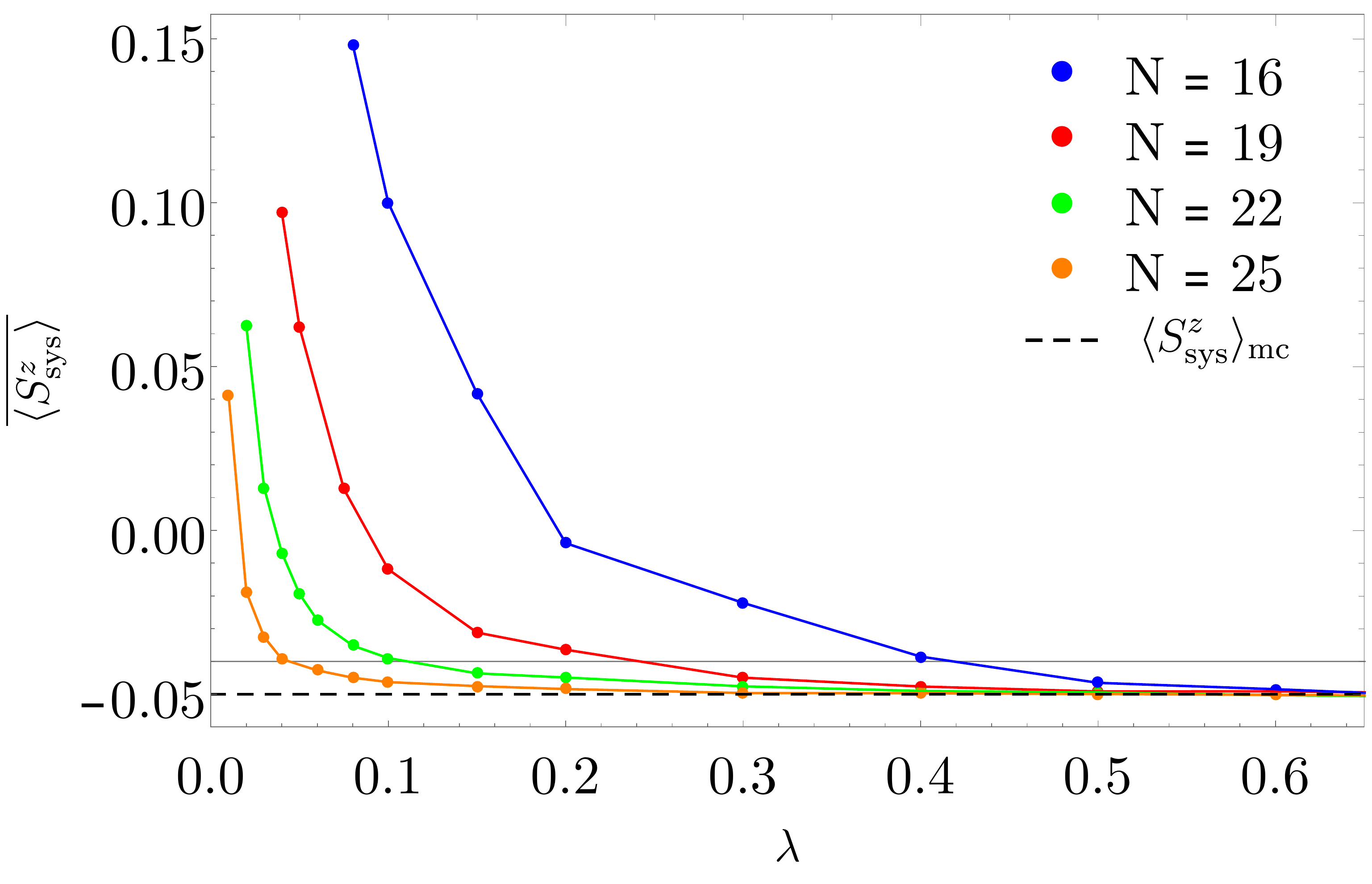}
	\caption{Longtime average value of the magnetization plotted over the interaction strength $\lambda$ for various bath sizes. 
	For sufficiently strong coupling the system thermalizes for all bath sizes. For sufficiently weak coupling the magnetization gets stuck for all bath sizes.}
	\label{fig3}
\end{figure}
\noindent
iii. {{\em superweak coupling regime:}   For very weak coupling (e.g. $\lambda = 0.01$) the magnetization does not decay to the thermal equilibrium value at all,
it rather gets stuck at a value closer to the initial value, which indicates the breakdown of FGR. This value depends on the interaction strength and on the bath size.
In accord with standard open quantum system theory, our below results indicate that this regime 
only exists for finite baths. We are not aware of any systematic approach to this regime in the literature to date. As the conflict between the GPB and FGR arises in the 
limit of weak interactions, cf. Eq. (\ref{scalnum}),  we are primarily interested in the transition from the Markovian to the superweak regime. 
A prime indicator of  superweak dynamics is, 
as mentioned above, the fact that $\langle S^z_{\text{sys}}(t) \rangle$ no longer decays down to the microcanonical expectation value 
$\langle S^z_{\text{sys}}\rangle_\text{mc}= -0.05$ as it does in the non-Markovian and the Markovian regime.\\

\noindent
Fig. \ref{fig3} shows the longtime average value of the magnetization plotted over the interaction strength $\lambda$ for various bath sizes.
For sufficiently strong coupling the magnetization decays to the thermal equilibrium value for all bath sizes. For each bath size there exists a critical
interaction strength $\lambda_{\text{crit}}$, below which the magnetization gets stuck at a non-thermal longtime average value, thus signaling the 
transition from the Markovian to the 
superweak regime.
This critical interaction strength $\lambda_{\text{crit}}$ decreases with bath size. We chose $\overline{\langle S^z_{\text{sys}}\rangle} = -0.04$ (horizontal grey line)
to define $\lambda_{\text{crit}}$. It turns out that the below scaling of $\lambda_{\text{crit}}$ is rather insensitive to the   exact positioning of this threshold, 
as long as it is sufficiently close to the thermal equilibrium value. Obviously, one expects $\overline{\langle S^z_{\text{sys}}\rangle} \rightarrow 0.5$ as $\lambda \rightarrow 0$.
\begin{figure}[H]
	\centering
  \includegraphics[width=0.48\textwidth]{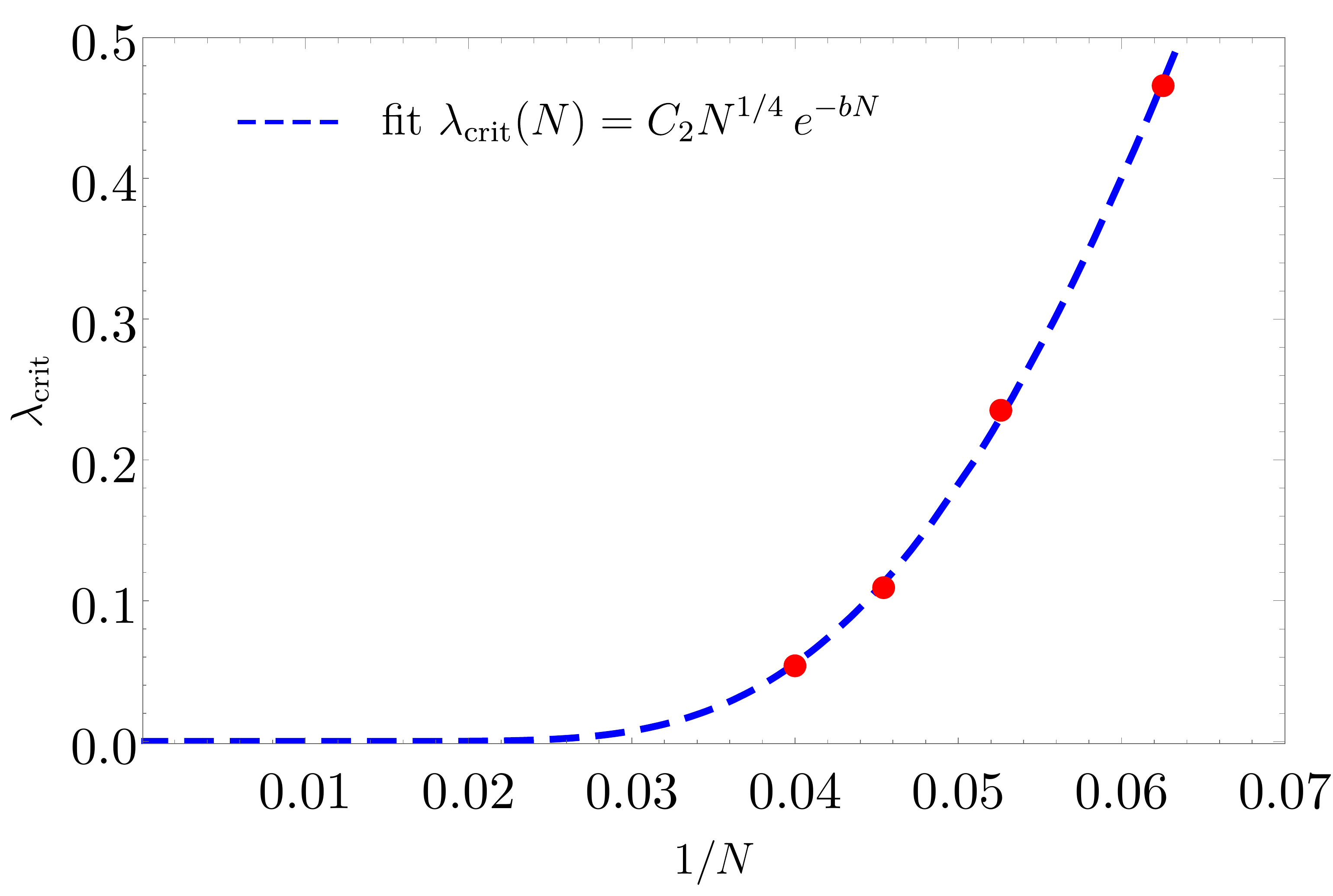}
	\caption{Critical interaction strength plotted over inverse system size. 
	The data is fitted as $\lambda_{\text{crit}}(N) = C_2 N^{1/4}\, \text{exp}^{-b N}$ with fit parameters $C_2 = 12.7$ and $b=0.25$. The principal form of this fit is
	motivated in Sect. \ref{scaltheo}.
	}
	\label{fig4}
\end{figure}
\noindent
Fig. \ref{fig4} displays the critical interaction strength plotted over the inverse system size. It strongly suggests that $\lambda_{\text{crit}} \rightarrow 0$ very quickly with 
increasing bath size $N$. Hence, for
all mesoscopic to macroscopic systems, and even more so in the thermodynamic limit, the transition to the superweak regime practically never occurs, such that behavior 
other than Markovian can hardly be expected even for physically very weak interactions. While this finding is another  main result of the quantitative analysis
at hand, it qualitatively hardly comes as a surprise in a larger context, given the practical success of Markovian quantum master equations. However, to elaborate on this result 
somewhat further, we present a theory that captures the data in Fig. \ref{fig4} rather accurately in Sect. \ref{scaltheo}.\newline

\noindent
Next we confirm the validity of FGR in the Markovian regime and discuss relaxation/equilibration times in all regimes. The motivation for the latter is twofold: 
On the one hand equilibration times enter the GPB (cf. Eq. (\ref{gpb})), on the other hand the 
scaling of equilibration times with the interaction strength may serve as a additional, quantitative indicator for the validity of FGR. Fig. \ref{fig5} displays
the observable dynamics $\langle S^z_{\text{sys}}(t) \rangle$ for $11$ randomly selected interaction strengths and bath sizes from the Markovian regime, which is lower bounded by  $\lambda_{\text{crit}}$ as determined from Fig. \ref{fig4} and upper bounded by $\lambda_{\text{non-Mark}} \approx 0.3$ for all $N$. Note the the time axis is scaled with the squared interaction strength
such that a collapse of the data onto one decaying exponential indicates the accordance with Eq. (\ref{decay}) and hence FGR. This collapse is evident.
\begin{figure}[H]
	\centering
	\includegraphics[width=0.48\textwidth]{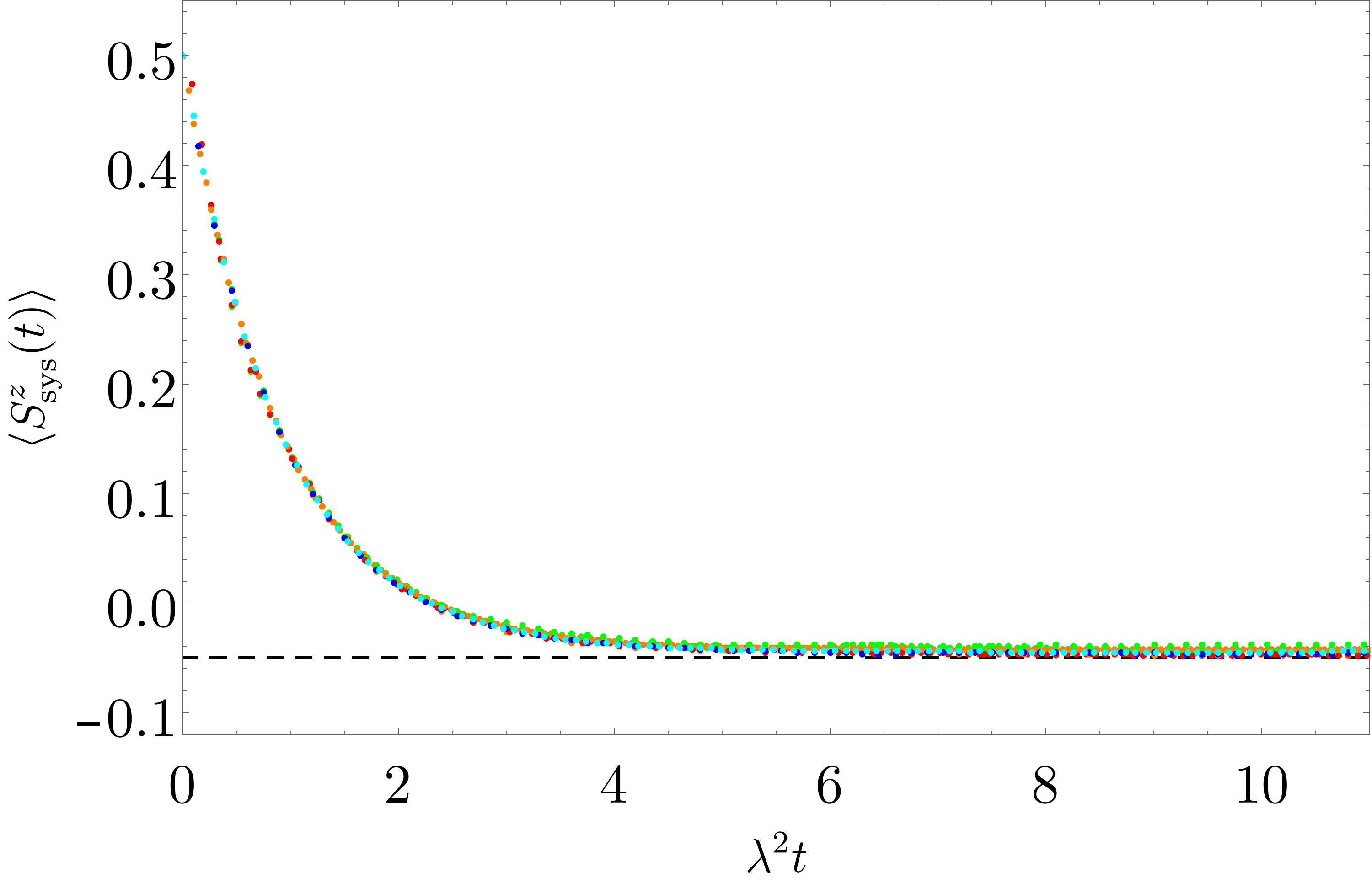}
	\caption{Exponential decays for various bath sizes and interaction strengths from the Markovian regime ($11$ curves randomly colored).}
	\label{fig5}
\end{figure}
\noindent
In Fig. \ref{fig6} the relaxation time $\tau_{\text{rel}}$ is plotted over $\lambda^{-2}$.
Here $\tau_{\text{rel}}$ is the time at which
the magnetization has decayed to $1/e$ of
its original value relative to the equilibrium value (cf. Eq. (\ref{decay})). 
%Given the essentially  monotonous dynamics of the present examples, this definition coincides with the one 
%given in Sect. \ref{garferm}. 
In the Markovian regime, i.e. for  
$ \lambda_{\text{non-Mark}}^{-2} \leq     \lambda^{-2}  \leq  \lambda^{-2}_{\text{crit}}   $, the relaxation time scales as 
$\tau_{\text{rel}} \sim \lambda^{-2\vphantom{\hat{2}}}$, as predicted by FGR, which also  confirms the applicability of FGR in the Markovian regime. 
At very small $\lambda$, i.e. in the 
superweak coupling regime, $\tau_{\text{rel}}$ first increases more slowly with increasing $\lambda^{-2}$ and likely  eventually even decreases to zero. From the data displayed
in Fig. \ref{fig6} this behavior is, however, qualitatively only visible for $N = 16$ due to numerical limitations at extremely small $\lambda$.
\begin{figure}[H]
	\centering
  \includegraphics[width=0.48\textwidth]{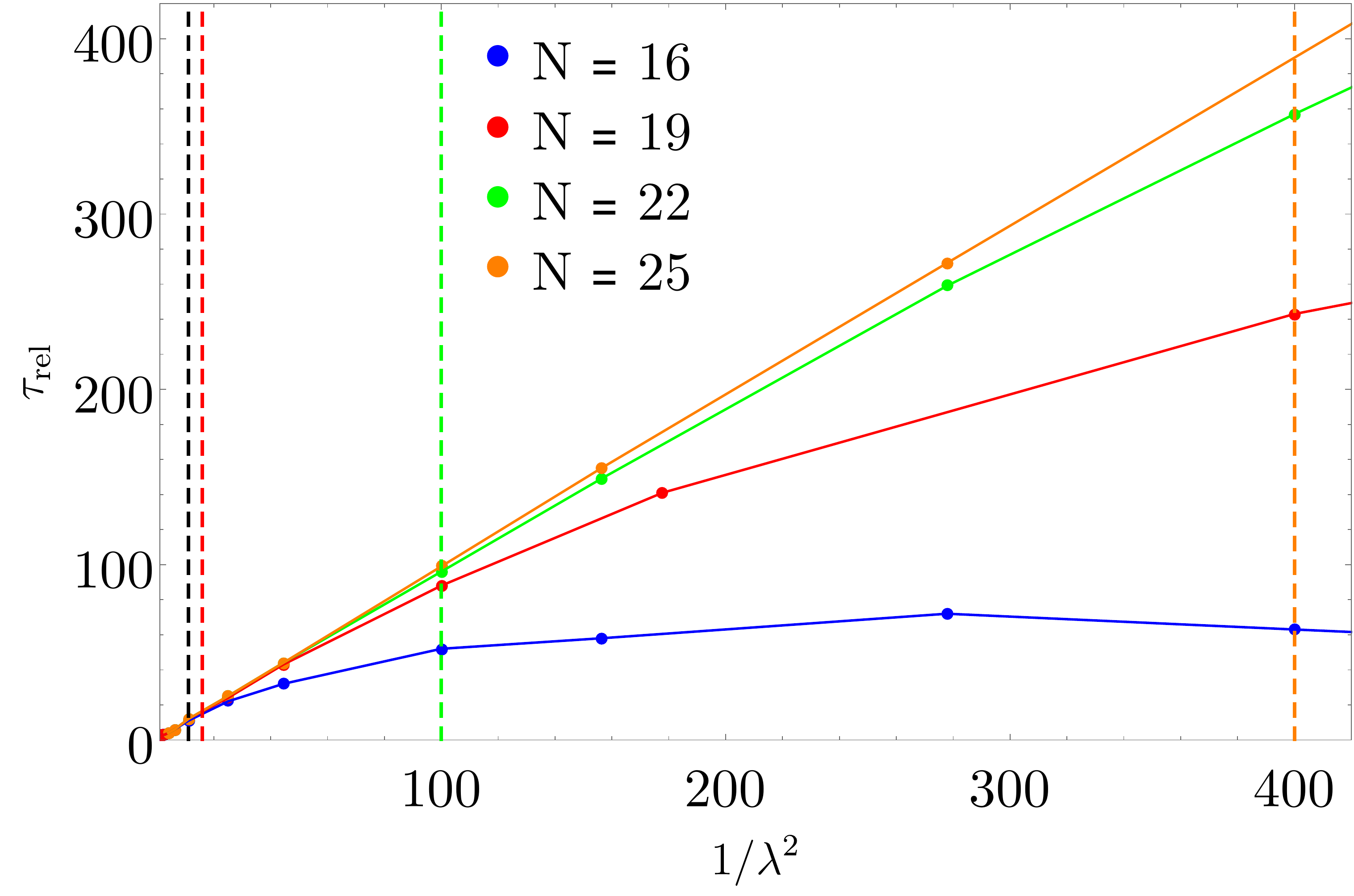}
	\caption{Relaxation time plotted over the inverse interaction strength squared for various bath sizes. Left of the vertical dashed black line lies the non-Markovian regime. Between the vertical dashed black line and the vertical dashed colored lines lies the respective Markovian regime, 
	i.e, the vertical dashed colored lines indicate the corresponding $\lambda^{-2}_{\text{crit}}$'s. For $N=16$ the Markovian regime does not exist.
	Within the  respective Markovian regimes $\tau_{\text{rel}} \approx 0.95 \lambda^{-2}$ holds for all system sizes in accord with FGR.} 
	\label{fig6}
\end{figure}
\newpage
\begin{figure}[H]
	\centering
	\includegraphics[width=0.48\textwidth]{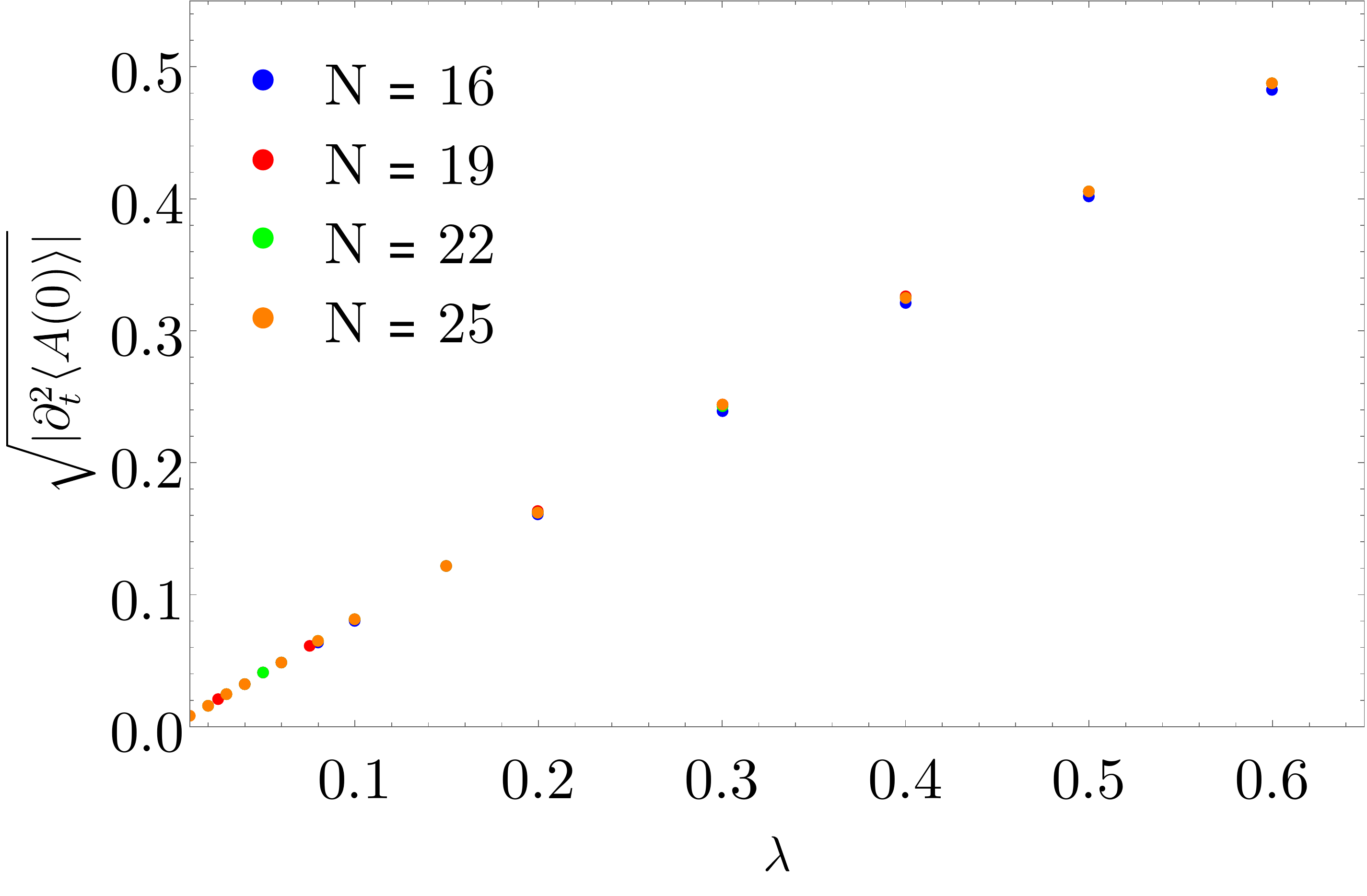}
	\caption{Square root of the initial curvature of the observable dynamics at $t=0$ plotted over the interaction strength. The data indicate that this quantity is 
	independent of the system size in all regimes of the interaction strength.}
	\label{fig7}
\end{figure}
\noindent
Eventually, we directly numerically  probe the connection between short time and long time dynamics suggested in Eq. \eqref{gpb}.  
To this end we compute the ``initial curvatures'' $\sqrt{|\partial_t^2 \langle A(t) \rangle\big{|}_{t=0}|}$ for various interaction strengths and 
systems sizes. The result is displayed in Fig. \ref{fig7}. 
As expected from Eq. (\ref{scalden}), and in full accord with a corresponding statement in Ref. \cite{garcia-pintos17}, the square root of the initial curvature scales linearly
with $\lambda$ and is practically independent of the system size. We are now set to assess the crucial numerator from Eq. \eqref{gpb} numerically.
From Eqs. (\ref{gpb}, \ref{times}) follows
\begin{eqnarray}
    \label{num}
    \pi a ||A||^{1/2} Q^{5/2} & =&  T_{\text{eq}} \sqrt{\bigg{|}\frac{\text{d}^2}{\text{d}t^2}\langle A(t) \rangle\big{|}_{t=0}\bigg{|}} \nonumber\\
    &\geq& \tau_{\text{rel}} \sqrt{\bigg{|}\frac{\text{d}^2}{\text{d}t^2}\langle A(t) \rangle\big{|}_{t=0}\bigg{|}} \,.
\end{eqnarray}
The lower bound to the numerator is displayed in Fig. \ref{fig8}. Recall that for a conclusive application of the GPB this numerator must be 
appropriately upper bounded. Correspondingly, Ref. \cite{garcia-pintos17} offers estimates for both $a$ and $Q$. While $a \sim  1$ is simply  traced back to the 
unimodality of $w$, the discussion on the order of magnitude of  $Q$ is quite involved. However, in the case of weak interactions, a microcanonical 
initial bath state comprising a large number of energy eigenstates, and an exponentially growing 
density of states in  the bath (with an exponent $\beta$ which is not too large),   $Q$ may also be expected to be of order unity, according to  
Ref. \cite{garcia-pintos17}. All these conditions apply to the models at hand. However, quite in contrast we find that the numerator grows at least up 
to values of ca. $55$ already for $N=25$ and interactions within the range of our numerical accessibility.  
Moreover, the data do not indicate any ``nearby'' upper bound of the numerator at $55$. This is at odds with a conclusive application of the GPB and 
another main result of the present paper.
\begin{figure}[H]
	\centering
	\includegraphics[width=0.48\textwidth]{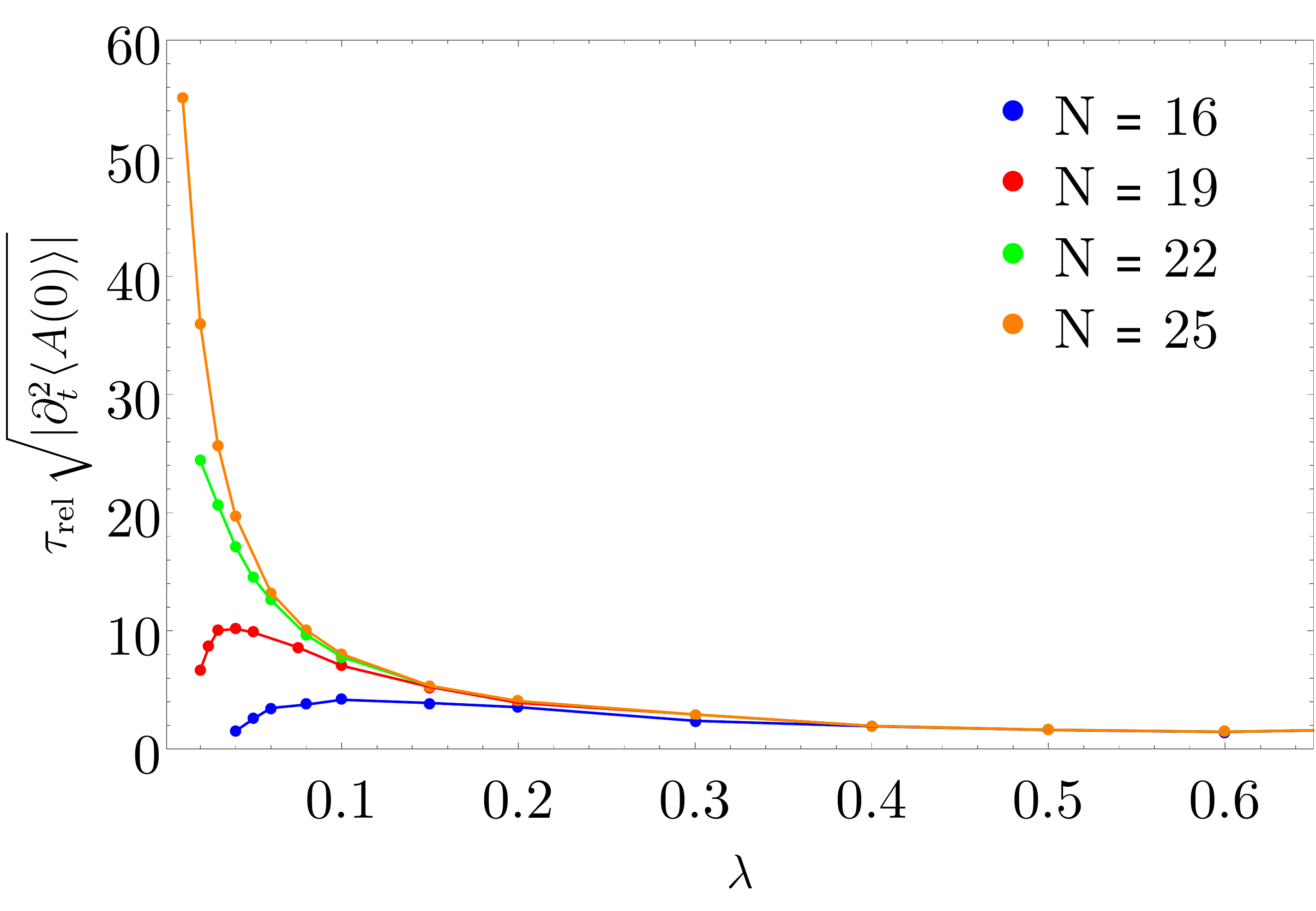}
	\caption{Numerator of Eq. (\ref{gpb}), i.e, the central quantity of the GPB plotted
	over the interaction strength. The numerator reaches values substantially larger than unity. The increase of the numerator with decreasing $\lambda$ extends into 
	the superweak regime.}
	\label{fig8}
\end{figure}
\noindent
Large numerators must occur, as outlined in Sect. \ref{garferm}, for large systems in the Markovian regime on the verge to the superweak regime. 
Although 
Fig. \ref{fig8} indicates that in the outer mathematical limit (which may be considered to be physically less relevant) $\lambda \rightarrow 0$, the 
numerator may eventually be of order unity, its growth appears to continue substantially into the superweak regime. While we are unable to verify this directly,
it appears plausible that the unbound growth of the numerator is due to an unbound growth of $a$, while $Q \sim 1$  may very well hold. We argue for this plausibility in 
Sect. \ref{discussion}.      \\
\noindent
We sum up this section as follows: In a potentially wide regime of interaction strengths, which is lower bounded by $\lambda_{\text{crit}}$, FGR is found to 
apply as suggested by standard open quantum system theory in the Markovian regime. The lower bound appears to decrease rapidly with system size $N$ (cf.
also Sect. \ref{scaltheo}), making it practically irrelevant for mesoscopic and macroscopic systems. 
This relates to the GPR inasmuch as the validity of FGR implies the breakdown of the practical applicability of the GPB at sufficiently  weak interactions.
We  numerically confirmed the occurrence of this breakdown directly for a system comprising $N=25$ spins.
%\begin{figure}[H]
%	\centering
	%\includegraphics[width=0.42\textwidth]{rand}
	%\caption{Decay of the magnetization for $L=8$ and $\lambda=0.2$. The bath couplings were drawn from a Gaussian distribution with mean one and standard deviation $0.2$. This breaks the rotational invariance of the bath (the composite system does not possess this symmetry due to the placement of the single spin). The decay looks almost exactly the same, suggesting a generic nature of the investigated system and stability versus small pertubations.}
	%\label{fig7}
%\end{figure}
%There is no hint that Fermi's Golden Rule breaks down for larger baths. These findings do not depend on the width of the energy window.
%\newline
%description of models, observables and ''upscaling``\\
%Results: No hint that FGR ever brakes down for infinitely large baths, FGR does break down for finite baths (breakdown-tiemscale related to DOS of the bath?), 
%equilibration time does scale with system size beyond the breakdown-timescale
%VERSCHIEDENE BAD ANFANGSZUSTAENDE\\
%KOENNEN WIR KLAR MACHEN DASS NICHT Q (SONDERN a) WAECHST?
%\newpage
\section{General scaling of $\lambda_\text{crit}$ with total system size}
\label{scaltheo}
While the  results on $\lambda_\text{crit}$ in Fig. \ref{fig4} are model dependent, a similar scaling may be expected whenever $H_\text{int}$ complies with the 
eigenstate thermalization hypothesis. This claim is substantiated in the following.
The starting point is the assumption that within the superweak regime, the overlap of the eigenstates of the uncoupled system $|n, \text{sys}+ \text{bath}  \rangle$ and 
those of the full system $|n,\text{sys}+ \text{bath}+ \text{int}\rangle$ is relatively large, i.e.
\begin{equation}
\label{stoer}
 |\langle n, \text{sys}+ \text{bath} |n,\text{sys}+ \text{bath}+ \text{int}\rangle| \approx 1\,. 
\end{equation}
A strong indication for this to occur arises from the leading order contributions to a perturbative correction to the eigenstates being small.
This condition may, according to textbook level perturbation theory, be approximated as
\begin{equation}
\label{stoer0}
 \lambda^2 \sum_{m\neq n}\frac{  |\langle m, \text{sys}+ \text{bath}|H_{\text{int}} |n,\text{sys}+ \text{bath}\rangle |^2 }{((m-n) \frac{1}{\Omega(N,\beta)})^2} \ll 1\,,
\end{equation}
where $\Omega(N,\beta)$ is the density of states of a system comprising $N$ spins (or other similar subsystems) at the energy that corresponds to the 
inverse temperature $\beta$. In Eq. (\ref{stoer0})   it is assumed that the level spacing within the relevant energy regime may be approximated as being constant.  
In this case the ``mean'' level spacing
is given by $1/\Omega(N,\beta)$. Following the eigenstate thermalization hypothesis ansatz \cite{6} we furthermore assume that there exists a typical value 
for the absolute squares of 
the matrix elements of the coupling operator, which varies with the energies $E_n, E_m$ only on energy scales much larger than the one relevant here. Furthermore, the 
eigenstate thermalization hypothesis suggests a specific scaling of these matrix elements with the density of states.
Following the eigenstate thermalization hypothesis we thus assume 
\begin{equation}
\label{eth}
  |\langle m, \text{sys}+ \text{bath}|H_{\text{int}} |n,\text{sys}+ \text{bath}\rangle |^2 \approx \frac{C_1}{\Omega(N,\beta)}\,,
\end{equation}
where $C_1$ is some real constant.
Exploiting this, Eq. (\ref{stoer0}) turns into
\begin{equation}
\label{stoer00}
 \lambda^2   C_1 \Omega(N,\beta)  \sum_{k\neq 0}\frac{ 1}{k^2} \ll 1\,.
\end{equation}
As the sum assumes the finite value $\pi^2/3$, we conclude
\begin{equation}
\label{crit}
\lambda_{\text{crit}} = \frac{C_2}{\sqrt{\Omega(N,\beta)}}
\end{equation}
for the scaling of  $\lambda_{\text{crit}}$, where $C_2$ is a constant whose concrete value depends on $ H_\text{sys}, H_\text{bath}$ and $H_\text{int} $. Now we turn to  an estimate for $\Omega(N,\beta)$.
For a sufficiently large Heisenberg spin system (or any other system consisting of $N$ similar, similarly and locally interacting subsystems)
it is reasonable to assume that the density of states is Gaussian with mean zero and variance proportional 
to the particle number $N$.
\begin{equation}
\Omega(N,E) \sim \dfrac{2^N}{\sqrt{N}} e^{-E^2/\alpha N}
\end{equation}
Using $\beta = \partial_E \log \Omega $ leads to 
\begin{equation}
\Omega(N,\beta) \sim \dfrac{1}{\sqrt{N}} e^{(\log 2 - 0.25 \alpha \beta^2) N}\,.
\end{equation}
Inserting this into Eq. (\ref{crit}) and fitting for $C_2$ and $\alpha$ yields the dashed line in Fig. \ref{fig4},
which matches the data quite well. This  remarkable agreement in turn backs up the argumentation which lead to Eq. (\ref{crit}). As the density of 
states is routinely expected to scale exponentially with the size of the system, Eq. (\ref{crit}) indicates that  $\lambda_{\text{crit}}$ will generally be exponentially 
small in the system size and thus the superweak regime will practically never be observed. 
\section{Garc\'{\i}a-Pintos bound and exponentially decaying observables }
\label{discussion}
As explained in Sect. \ref{garferm}
the applicability of the GPB hinges on the two parameters $a$ and  $Q$, both of which  should be of order unity to establish a meaningful relation between the short time dynamics and the equilibration time 
in the sense of the GPB. However, for some of the numerical examples considered in Sect. \ref{spinexp} at least one of the parameters must be substantially larger than
unity. While we are unable to perform a direct numerical check for large system sizes, we strongly suspect that $a\sim 1$ is 
violated at weak interactions even though the corresponding $w(G)$ (cf. Sect. \ref{garferm}) is strictly unimodal. In the remainder of the present section we
explain and back up this claim.\\
\noindent
Consider the mathematically simple case of an infinite temperature environment in the initial state
\mbox{$\rho =(S^z_\text{sys}+1_\text{sys}/2) \otimes 1_\text{bath}/d_\text{bath}$.} This initial state yields 
\begin{equation}
\langle S^z_\text{sys}(t) \rangle = \text{Tr}\{  S^z_\text{sys}(t) \rho   \}  =  \text{Tr}\{  S^z_\text{sys}(t) S^z_\text{sys}  \}                     
\end{equation}
for the dynamics of the observable 
$\langle S^z_\text{sys}(t) \rangle$, which may be rewritten as 
\begin{equation}
\label{dynfour}
\langle S^z_\text{sys}(t) \rangle = \sum_{j,k}|\langle j| S^z_\text{sys}|k \rangle |^2 e^{\mathrm{i}(E_j-E_k)t}\,.                    
\end{equation}
Now consider the distribution $p_{jk}$ as defined in Eq. (\ref{palfa}) for this initial state and choice of observable, i.e., $A=S^z_\text{sys}$.
\begin{equation}
\label{palfa1}
    p_{jk} \propto  |\langle j| S^z_\text{sys}|k \rangle ^2 | =  |\langle j| S^z_\text{sys}|k \rangle |^2 \quad \text{for} \quad j\neq k            
\end{equation}
Due to the system being non-integrable in the sense of a Bethe ansatz, the eigenstate thermalization 
hypothesis may be expected to hold, yielding  $|\langle j| S^z_\text{sys}|j \rangle |^2 \approx 0$. Exploiting this, the insertion of Eq. (\ref{palfa1}) into
Eq. (\ref{dynfour}) yields
\begin{equation}
    \label{dynfour1}
    \langle S^z_\text{sys}(t) \rangle \propto \sum_{j,k} p_{jk}  e^{\mathrm{i}(E_j-E_k)t}\,.                    
    \end{equation}
To the extend to which  $p_{jk}$ may indeed be replaced by a smooth probability density as discussed around Eq. (\ref{wege}), Eq. (\ref{dynfour1}) may be 
rewritten as 
\begin{equation}
    \label{dynfour2}
    \langle S^z_\text{sys}(t) \rangle \propto \int w(G) e^{\mathrm{i}Gt} \text{d} G                   \,.
    \end{equation}
Thus, for the present scenario, $w(G)$ is essentially the Fourier transform of the observable dynamics $\langle S^z_\text{sys}(t) \rangle$. Based on the 
numerical findings displayed, e.g.,  in Fig. \ref{fig2} it appears plausible that $\langle S^z_\text{sys}(t) \rangle$ will be  an exponential decay for
infinite temperature initial states as well. Therefore, $w(G)$ will be Lorentzian. While a Lorentzian distribution is clearly unimodal with one well-behaved maximum,
 its variance diverges. Consequently $a$, as defined in Eq. (\ref{defaq}), diverges as well. Thus, in contrast to the assumptions in 
 Ref. \cite{garcia-pintos17}, $a\sim 1$ does not hold. This is the last main result of the present paper. Of course $\sigma_G$ cannot really diverge in any system featuring a finite energy spectrum. 
 However, the finiteness of the spectrum essentially causes a cut-off of the tails of the Lorentzian at some frequency. This cut-off actually renders   
the standard deviation $\sigma_G$ finite. Nonetheless, this standard deviation does not reasonably reflect the width of $w(G)$. It will be much larger than
other measures of the width such as the full-width-at-half-maximum, etc. \\
\noindent
Some attention should also be paid to the question whether or not the GPB scales with the size of the environment. (Earlier works presented
upper bounds that explicitly depend on the size of the environment, which is often seen as a drawback \cite{goldstein, malabarba}). While the GPB does not explicitly dependent on the size of 
the 
environment, the latter may enter via the parameter $a$. For any (weak) interaction strength $\lambda$ there exists a system size $N (\lambda_{\text{crit}})$ above which 
FGR applies.
Above that size the GPB is independent of $N$. Below or at  $N (\lambda_{\text{crit}})$, however, the numerator in Eq. (\ref{gpb}) may depend 
on $N$ rather strongly. For arbitrarily small $\lambda$ this $N (\lambda_{\text{crit}})$ may become arbitrarily large. Thus,
in the class of models discussed in the paper at hand, one can always  find instances for which the GPB depends on system size even for very large systems, i.e. $N \gg 1$.

%This appears to hold true for a wide range of initial states, including states with a high Von Neumann entropy of the bath.

%in genral FGR wins\\

%the main fallacy of the GP-bound is in the assumptions on a: a may grow beyond any bound even if p is unimodal, namely if p is of Lorentzian type. 
%a may scale with bath size beyond the breakdown timescale\\

%p is closely related to Fourier transform of observable dynamics $->$ p likely to be Lorentzian for any exponential or exponentially damped dynamics.

\section{Summary and conclusion}
\label{sumcon}
In the paper at hand we conceptually and numerically analyzed an upper bound on equilibration times presented  in a recent paper by Garc\'{\i}a-Pintos et al.
To this end, we investigated a standard system-bath setup by monitoring the system's magnetization for various bath sizes and interaction strengths. 
This numerical investigation is based on the solution of the time dependent Schr\"odinger equation for the full system, including the bath.
We identified a Markovian regime of interaction strengths $\lambda$  in which Fermi's Golden Rule holds, i.e., the system thermalizes in an exponential way
and the equilibration time scales as $\lambda^{-2}$. This relates to the Garc\'{\i}a-Pintos bound inasmuch as the validity of Fermi's Golden Rule and the usefulness of the Garc\'{\i}a-Pintos bound are analytically shown 
to be mutually exclusive at sufficiently 
small $\lambda$. At extremely small  $\lambda$,  we indeed find a ``superweak'' regime in which Fermi's Golden rule does not apply. This regime (in principle) 
exists for 
finite baths  and 
is reached below some  $\lambda_{\text{crit}}$ which is shown to scale inversely exponentially in the bath size, suggesting that the superweak regime  practically ceases to exist when considering moderately large systems.
However, in  the superweak regime the Garc\'{\i}a-Pintos bound may eventullay regain applicability, 
although its non-applicability is found to extend also into the superweak regime.
\begin{acknowledgments}
Stimulating discussions with R. Steinigeweg and J. Richter are gratefully acknowledged. We also thank L. P. Garc\'{\i}a-Pintos and A. M. Alhambra for interesting discussions, as well as A. J. Short for a comment on an early version of this paper.
This work was supported by the Deutsche Forschungsgemeinschaft (DFG) within the Research Unit FOR 2692 under Grant No. 397107022.
\end{acknowledgments}
\vspace*{-5px}
\bibliography{literature}

\begin{thebibliography}{33}%
\makeatletter
\providecommand \@ifxundefined [1]{%
 \@ifx{#1\undefined}
}%
\providecommand \@ifnum [1]{%
 \ifnum #1\expandafter \@firstoftwo
 \else \expandafter \@secondoftwo
 \fi
}%
\providecommand \@ifx [1]{%
 \ifx #1\expandafter \@firstoftwo
 \else \expandafter \@secondoftwo
 \fi
}%
\providecommand \natexlab [1]{#1}%
\providecommand \enquote  [1]{``#1''}%
\providecommand \bibnamefont  [1]{#1}%
\providecommand \bibfnamefont [1]{#1}%
\providecommand \citenamefont [1]{#1}%
\providecommand \href@noop [0]{\@secondoftwo}%
\providecommand \href [0]{\begingroup \@sanitize@url \@href}%
\providecommand \@href[1]{\@@startlink{#1}\@@href}%
\providecommand \@@href[1]{\endgroup#1\@@endlink}%
\providecommand \@sanitize@url [0]{\catcode `\\12\catcode `\$12\catcode
  `\&12\catcode `\#12\catcode `\^12\catcode `\_12\catcode `\%12\relax}%
\providecommand \@@startlink[1]{}%
\providecommand \@@endlink[0]{}%
\providecommand \url  [0]{\begingroup\@sanitize@url \@url }%
\providecommand \@url [1]{\endgroup\@href {#1}{\urlprefix }}%
\providecommand \urlprefix  [0]{URL }%
\providecommand \Eprint [0]{\href }%
\providecommand \doibase [0]{http://dx.doi.org/}%
\providecommand \selectlanguage [0]{\@gobble}%
\providecommand \bibinfo  [0]{\@secondoftwo}%
\providecommand \bibfield  [0]{\@secondoftwo}%
\providecommand \translation [1]{[#1]}%
\providecommand \BibitemOpen [0]{}%
\providecommand \bibitemStop [0]{}%
\providecommand \bibitemNoStop [0]{.\EOS\space}%
\providecommand \EOS [0]{\spacefactor3000\relax}%
\providecommand \BibitemShut  [1]{\csname bibitem#1\endcsname}%
\let\auto@bib@innerbib\@empty
%</preamble>
\bibitem [{\citenamefont {Gogolin}\ and\ \citenamefont
  {Eisert}(2016)}]{gogolin16}%
  \BibitemOpen
  \bibfield  {author} {\bibinfo {author} {\bibfnamefont {C.}~\bibnamefont
  {Gogolin}}\ and\ \bibinfo {author} {\bibfnamefont {J.}~\bibnamefont
  {Eisert}},\ }\bibfield  {title} {\enquote {\bibinfo {title} {Equilibration,
  thermalisation, and the emergence of statistical mechanics in closed quantum
  systems},}\ }\href {\doibase 10.1088/0034-4885/79/5/056001} {\bibfield
  {journal} {\bibinfo  {journal} {Reports on Progress in Physics}\ }\textbf
  {\bibinfo {volume} {79}},\ \bibinfo {pages} {056001} (\bibinfo {year}
  {2016})}\BibitemShut {NoStop}%
\bibitem [{\citenamefont {Goldstein}\ \emph {et~al.}(2006)\citenamefont
  {Goldstein}, \citenamefont {Lebowitz}, \citenamefont {Tumulka},\ and\
  \citenamefont {Zangh\`{\i}}}]{cantyp}%
  \BibitemOpen
  \bibfield  {author} {\bibinfo {author} {\bibfnamefont {S.}~\bibnamefont
  {Goldstein}}, \bibinfo {author} {\bibfnamefont {J.~L.}\ \bibnamefont
  {Lebowitz}}, \bibinfo {author} {\bibfnamefont {R.}~\bibnamefont {Tumulka}}, \
  and\ \bibinfo {author} {\bibfnamefont {N.}~\bibnamefont {Zangh\`{\i}}},\
  }\bibfield  {title} {\enquote {\bibinfo {title} {Canonical typicality},}\
  }\href {\doibase 10.1103/PhysRevLett.96.050403} {\bibfield  {journal}
  {\bibinfo  {journal} {Phys. Rev. Lett.}\ }\textbf {\bibinfo {volume} {96}},\
  \bibinfo {pages} {050403} (\bibinfo {year} {2006})}\BibitemShut {NoStop}%
\bibitem [{\citenamefont {Reimann}(2007)}]{reimann}%
  \BibitemOpen
  \bibfield  {author} {\bibinfo {author} {\bibfnamefont {P.}~\bibnamefont
  {Reimann}},\ }\bibfield  {title} {\enquote {\bibinfo {title} {Typicality for
  generalized microcanonical ensembles},}\ }\href {\doibase
  10.1103/PhysRevLett.99.160404} {\bibfield  {journal} {\bibinfo  {journal}
  {Phys. Rev. Lett.}\ }\textbf {\bibinfo {volume} {99}},\ \bibinfo {pages}
  {160404} (\bibinfo {year} {2007})}\BibitemShut {NoStop}%
\bibitem [{\citenamefont {Lloyd}(1988)}]{lloyd}%
  \BibitemOpen
  \bibfield  {author} {\bibinfo {author} {\bibfnamefont {S.}~\bibnamefont
  {Lloyd}},\ }\bibfield  {title} {\enquote {\bibinfo {title} {Pure state
  quantum statistical mechanics and black holes},}\ }\href
  {https://arxiv.org/abs/1307.0378} {\bibfield  {journal} {\bibinfo  {journal}
  {arXiv:1307.0378v1}\ } (\bibinfo {year} {1988})}\BibitemShut {NoStop}%
\bibitem [{\citenamefont {Srednicki}(1994)}]{schrecknicki}%
  \BibitemOpen
  \bibfield  {author} {\bibinfo {author} {\bibfnamefont {M.}~\bibnamefont
  {Srednicki}},\ }\bibfield  {title} {\enquote {\bibinfo {title} {Chaos and
  quantum thermalization},}\ }\href {\doibase 10.1103/PhysRevE.50.888}
  {\bibfield  {journal} {\bibinfo  {journal} {Phys. Rev. E}\ }\textbf {\bibinfo
  {volume} {50}},\ \bibinfo {pages} {888--901} (\bibinfo {year}
  {1994})}\BibitemShut {NoStop}%
\bibitem [{\citenamefont {Deutsch}(1991)}]{deutsch}%
  \BibitemOpen
  \bibfield  {author} {\bibinfo {author} {\bibfnamefont {J.~M.}\ \bibnamefont
  {Deutsch}},\ }\bibfield  {title} {\enquote {\bibinfo {title} {Quantum
  statistical mechanics in a closed system},}\ }\href {\doibase
  10.1103/PhysRevA.43.2046} {\bibfield  {journal} {\bibinfo  {journal} {Phys.
  Rev. A}\ }\textbf {\bibinfo {volume} {43}},\ \bibinfo {pages} {2046--2049}
  (\bibinfo {year} {1991})}\BibitemShut {NoStop}%
\bibitem [{\citenamefont {Reimann}(2008)}]{reimann2}%
  \BibitemOpen
  \bibfield  {author} {\bibinfo {author} {\bibfnamefont {P.}~\bibnamefont
  {Reimann}},\ }\bibfield  {title} {\enquote {\bibinfo {title} {Foundation of
  statistical mechanics under experimentally realistic conditions},}\ }\href
  {\doibase 10.1103/PhysRevLett.101.190403} {\bibfield  {journal} {\bibinfo
  {journal} {Phys. Rev. Lett.}\ }\textbf {\bibinfo {volume} {101}},\ \bibinfo
  {pages} {190403} (\bibinfo {year} {2008})}\BibitemShut {NoStop}%
\bibitem [{\citenamefont {Linden}\ \emph {et~al.}(2009)\citenamefont {Linden},
  \citenamefont {Popescu}, \citenamefont {Short},\ and\ \citenamefont
  {Winter}}]{linden}%
  \BibitemOpen
  \bibfield  {author} {\bibinfo {author} {\bibfnamefont {N.}~\bibnamefont
  {Linden}}, \bibinfo {author} {\bibfnamefont {S.}~\bibnamefont {Popescu}},
  \bibinfo {author} {\bibfnamefont {A.~J.}\ \bibnamefont {Short}}, \ and\
  \bibinfo {author} {\bibfnamefont {A.}~\bibnamefont {Winter}},\ }\bibfield
  {title} {\enquote {\bibinfo {title} {Quantum mechanical evolution towards
  thermal equilibrium},}\ }\href {\doibase 10.1103/PhysRevE.79.061103}
  {\bibfield  {journal} {\bibinfo  {journal} {Phys. Rev. E}\ }\textbf {\bibinfo
  {volume} {79}},\ \bibinfo {pages} {061103} (\bibinfo {year}
  {2009})}\BibitemShut {NoStop}%
\bibitem [{\citenamefont {Short}\ and\ \citenamefont {Farrelly}(2012)}]{4}%
  \BibitemOpen
  \bibfield  {author} {\bibinfo {author} {\bibfnamefont {A.~J.}\ \bibnamefont
  {Short}}\ and\ \bibinfo {author} {\bibfnamefont {T.~C.}\ \bibnamefont
  {Farrelly}},\ }\bibfield  {title} {\enquote {\bibinfo {title} {Quantum
  equilibration in finite time},}\ }\href {\doibase
  10.1088/1367-2630/14/1/013063} {\bibfield  {journal} {\bibinfo  {journal}
  {New Journal of Physics}\ }\textbf {\bibinfo {volume} {14}},\ \bibinfo
  {pages} {013063} (\bibinfo {year} {2012})}\BibitemShut {NoStop}%
\bibitem [{\citenamefont {Goldstein}\ \emph {et~al.}(2013)\citenamefont
  {Goldstein}, \citenamefont {Hara},\ and\ \citenamefont {Tasaki}}]{goldstein}%
  \BibitemOpen
  \bibfield  {author} {\bibinfo {author} {\bibfnamefont {S.}~\bibnamefont
  {Goldstein}}, \bibinfo {author} {\bibfnamefont {T.}~\bibnamefont {Hara}}, \
  and\ \bibinfo {author} {\bibfnamefont {H.}~\bibnamefont {Tasaki}},\
  }\bibfield  {title} {\enquote {\bibinfo {title} {Time scales in the approach
  to equilibrium of macroscopic quantum systems},}\ }\href {\doibase
  10.1103/PhysRevLett.111.140401} {\bibfield  {journal} {\bibinfo  {journal}
  {Phys. Rev. Lett.}\ }\textbf {\bibinfo {volume} {111}},\ \bibinfo {pages}
  {140401} (\bibinfo {year} {2013})}\BibitemShut {NoStop}%
\bibitem [{\citenamefont {Malabarba}\ \emph {et~al.}(2014)\citenamefont
  {Malabarba}, \citenamefont {Garc\'{\i}a-Pintos}, \citenamefont {Linden},
  \citenamefont {Farrelly},\ and\ \citenamefont {Short}}]{malabarba}%
  \BibitemOpen
  \bibfield  {author} {\bibinfo {author} {\bibfnamefont {A.~S.~L.}\
  \bibnamefont {Malabarba}}, \bibinfo {author} {\bibfnamefont {L.~P.}\
  \bibnamefont {Garc\'{\i}a-Pintos}}, \bibinfo {author} {\bibfnamefont
  {N.}~\bibnamefont {Linden}}, \bibinfo {author} {\bibfnamefont {T.~C.}\
  \bibnamefont {Farrelly}}, \ and\ \bibinfo {author} {\bibfnamefont {A.~J.}\
  \bibnamefont {Short}},\ }\bibfield  {title} {\enquote {\bibinfo {title}
  {Quantum systems equilibrate rapidly for most observables},}\ }\href
  {\doibase 10.1103/PhysRevE.90.012121} {\bibfield  {journal} {\bibinfo
  {journal} {Phys. Rev. E}\ }\textbf {\bibinfo {volume} {90}},\ \bibinfo
  {pages} {012121} (\bibinfo {year} {2014})}\BibitemShut {NoStop}%
\bibitem [{\citenamefont {Kastner}(2011)}]{kastner}%
  \BibitemOpen
  \bibfield  {author} {\bibinfo {author} {\bibfnamefont {M.}~\bibnamefont
  {Kastner}},\ }\bibfield  {title} {\enquote {\bibinfo {title} {Diverging
  equilibration times in long-range quantum spin models},}\ }\href {\doibase
  10.1103/PhysRevLett.106.130601} {\bibfield  {journal} {\bibinfo  {journal}
  {Phys. Rev. Lett.}\ }\textbf {\bibinfo {volume} {106}},\ \bibinfo {pages}
  {130601} (\bibinfo {year} {2011})}\BibitemShut {NoStop}%
\bibitem [{\citenamefont {Schiulaz}\ \emph {et~al.}(2019)\citenamefont
  {Schiulaz}, \citenamefont {Torres-Herrera},\ and\ \citenamefont
  {Santos}}]{5}%
  \BibitemOpen
  \bibfield  {author} {\bibinfo {author} {\bibfnamefont {M.}~\bibnamefont
  {Schiulaz}}, \bibinfo {author} {\bibfnamefont {E.~J.}\ \bibnamefont
  {Torres-Herrera}}, \ and\ \bibinfo {author} {\bibfnamefont {L.~F.}\
  \bibnamefont {Santos}},\ }\bibfield  {title} {\enquote {\bibinfo {title}
  {Thouless and relaxation time scales in many-body quantum systems},}\ }\href
  {\doibase 10.1103/PhysRevB.99.174313} {\bibfield  {journal} {\bibinfo
  {journal} {Phys. Rev. B}\ }\textbf {\bibinfo {volume} {99}},\ \bibinfo
  {pages} {174313} (\bibinfo {year} {2019})}\BibitemShut {NoStop}%
\bibitem [{\citenamefont {Cramer}\ \emph {et~al.}(2008)\citenamefont {Cramer},
  \citenamefont {Dawson}, \citenamefont {Eisert},\ and\ \citenamefont
  {Osborne}}]{cramer}%
  \BibitemOpen
  \bibfield  {author} {\bibinfo {author} {\bibfnamefont {M.}~\bibnamefont
  {Cramer}}, \bibinfo {author} {\bibfnamefont {C.~M.}\ \bibnamefont {Dawson}},
  \bibinfo {author} {\bibfnamefont {J.}~\bibnamefont {Eisert}}, \ and\ \bibinfo
  {author} {\bibfnamefont {T.~J.}\ \bibnamefont {Osborne}},\ }\bibfield
  {title} {\enquote {\bibinfo {title} {Exact relaxation in a class of
  nonequilibrium quantum lattice systems},}\ }\href {\doibase
  10.1103/PhysRevLett.100.030602} {\bibfield  {journal} {\bibinfo  {journal}
  {Phys. Rev. Lett.}\ }\textbf {\bibinfo {volume} {100}},\ \bibinfo {pages}
  {030602} (\bibinfo {year} {2008})}\BibitemShut {NoStop}%
\bibitem [{\citenamefont {Diez}\ \emph {et~al.}(2010)\citenamefont {Diez},
  \citenamefont {Chancellor}, \citenamefont {Haas}, \citenamefont {Venuti},\
  and\ \citenamefont {Zanardi}}]{diez}%
  \BibitemOpen
  \bibfield  {author} {\bibinfo {author} {\bibfnamefont {M.}~\bibnamefont
  {Diez}}, \bibinfo {author} {\bibfnamefont {N.}~\bibnamefont {Chancellor}},
  \bibinfo {author} {\bibfnamefont {S.}~\bibnamefont {Haas}}, \bibinfo {author}
  {\bibfnamefont {L.~C.}\ \bibnamefont {Venuti}}, \ and\ \bibinfo {author}
  {\bibfnamefont {P.}~\bibnamefont {Zanardi}},\ }\bibfield  {title} {\enquote
  {\bibinfo {title} {Local quenches in frustrated quantum spin chains: Global
  versus subsystem equilibration},}\ }\href {\doibase
  10.1103/PhysRevA.82.032113} {\bibfield  {journal} {\bibinfo  {journal} {Phys.
  Rev. A}\ }\textbf {\bibinfo {volume} {82}},\ \bibinfo {pages} {032113}
  (\bibinfo {year} {2010})}\BibitemShut {NoStop}%
\bibitem [{\citenamefont {Vinayak}\ and\ \citenamefont
  {{\v{Z}}nidari{\v{c}}}(2012)}]{vinayak}%
  \BibitemOpen
  \bibfield  {author} {\bibinfo {author} {\bibnamefont {Vinayak}}\ and\
  \bibinfo {author} {\bibfnamefont {M.}~\bibnamefont {{\v{Z}}nidari{\v{c}}}},\
  }\bibfield  {title} {\enquote {\bibinfo {title} {Subsystem dynamics under
  random hamiltonian evolution},}\ }\href {\doibase
  10.1088/1751-8113/45/12/125204} {\bibfield  {journal} {\bibinfo  {journal}
  {Journal of Physics A: Mathematical and Theoretical}\ }\textbf {\bibinfo
  {volume} {45}},\ \bibinfo {pages} {125204} (\bibinfo {year}
  {2012})}\BibitemShut {NoStop}%
\bibitem [{\citenamefont {Torres-Herrera}\ and\ \citenamefont
  {Santos}(2014)}]{torres}%
  \BibitemOpen
  \bibfield  {author} {\bibinfo {author} {\bibfnamefont {E.~J.}\ \bibnamefont
  {Torres-Herrera}}\ and\ \bibinfo {author} {\bibfnamefont {L.~F.}\
  \bibnamefont {Santos}},\ }\bibfield  {title} {\enquote {\bibinfo {title}
  {Quench dynamics of isolated many-body quantum systems},}\ }\href {\doibase
  10.1103/PhysRevA.89.043620} {\bibfield  {journal} {\bibinfo  {journal} {Phys.
  Rev. A}\ }\textbf {\bibinfo {volume} {89}},\ \bibinfo {pages} {043620}
  (\bibinfo {year} {2014})}\BibitemShut {NoStop}%
\bibitem [{\citenamefont {Garc\'{\i}a-Pintos}\ \emph
  {et~al.}(2017)\citenamefont {Garc\'{\i}a-Pintos}, \citenamefont {Linden},
  \citenamefont {Malabarba}, \citenamefont {Short},\ and\ \citenamefont
  {Winter}}]{garcia-pintos17}%
  \BibitemOpen
  \bibfield  {author} {\bibinfo {author} {\bibfnamefont {L.~P.}\ \bibnamefont
  {Garc\'{\i}a-Pintos}}, \bibinfo {author} {\bibfnamefont {N.}~\bibnamefont
  {Linden}}, \bibinfo {author} {\bibfnamefont {A.~S.~L.}\ \bibnamefont
  {Malabarba}}, \bibinfo {author} {\bibfnamefont {A.~J.}\ \bibnamefont
  {Short}}, \ and\ \bibinfo {author} {\bibfnamefont {A.}~\bibnamefont
  {Winter}},\ }\bibfield  {title} {\enquote {\bibinfo {title} {Equilibration
  time scales of physically relevant observables},}\ }\href {\doibase
  10.1103/PhysRevX.7.031027} {\bibfield  {journal} {\bibinfo  {journal} {Phys.
  Rev. X}\ }\textbf {\bibinfo {volume} {7}},\ \bibinfo {pages} {031027}
  (\bibinfo {year} {2017})}\BibitemShut {NoStop}%
\bibitem [{\citenamefont {Hove}(1957)}]{hove}%
  \BibitemOpen
  \bibfield  {author} {\bibinfo {author} {\bibfnamefont {L.~Van}\ \bibnamefont
  {Hove}},\ }\bibfield  {title} {\enquote {\bibinfo {title} {The approach to
  equilibrium in quantum statistics: A perturbation treatment to general
  order},}\ }\href
  {http://www.sciencedirect.com/science/article/pii/S0031891457928914}
  {\bibfield  {journal} {\bibinfo  {journal} {Physica}\ }\textbf {\bibinfo
  {volume} {23}},\ \bibinfo {pages} {441 -- 480} (\bibinfo {year}
  {1957})}\BibitemShut {NoStop}%
\bibitem [{\citenamefont {Bartsch}\ \emph {et~al.}(2008)\citenamefont
  {Bartsch}, \citenamefont {Steinigeweg},\ and\ \citenamefont
  {Gemmer}}]{bartsch}%
  \BibitemOpen
  \bibfield  {author} {\bibinfo {author} {\bibfnamefont {C.}~\bibnamefont
  {Bartsch}}, \bibinfo {author} {\bibfnamefont {R.}~\bibnamefont
  {Steinigeweg}}, \ and\ \bibinfo {author} {\bibfnamefont {J.}~\bibnamefont
  {Gemmer}},\ }\bibfield  {title} {\enquote {\bibinfo {title} {Occurrence of
  exponential relaxation in closed quantum systems},}\ }\href {\doibase
  10.1103/PhysRevE.77.011119} {\bibfield  {journal} {\bibinfo  {journal} {Phys.
  Rev. E}\ }\textbf {\bibinfo {volume} {77}},\ \bibinfo {pages} {011119}
  (\bibinfo {year} {2008})}\BibitemShut {NoStop}%
\bibitem [{\citenamefont {Joos}\ \emph {et~al.}(2003)\citenamefont {Joos},
  \citenamefont {Zeh}, \citenamefont {Kiefer}, \citenamefont {Giulini},
  \citenamefont {Kupsch},\ and\ \citenamefont {Stamatescu}}]{kupsch}%
  \BibitemOpen
  \bibfield  {author} {\bibinfo {author} {\bibfnamefont {E.}~\bibnamefont
  {Joos}}, \bibinfo {author} {\bibfnamefont {H.~D.}\ \bibnamefont {Zeh}},
  \bibinfo {author} {\bibfnamefont {C.}~\bibnamefont {Kiefer}}, \bibinfo
  {author} {\bibfnamefont {D.~J.~W.}\ \bibnamefont {Giulini}}, \bibinfo
  {author} {\bibfnamefont {J.}~\bibnamefont {Kupsch}}, \ and\ \bibinfo {author}
  {\bibfnamefont {I.~O.}\ \bibnamefont {Stamatescu}},\ }\href
  {https://books.google.de/books?id=6eTHcxeNxdUC} {\emph {\bibinfo {title}
  {Decoherence and the Appearance of a Classical World in Quantum Theory}}},\
  Physics and astronomy online library\ (\bibinfo  {publisher} {Springer},\
  \bibinfo {year} {2003})\BibitemShut {NoStop}%
\bibitem [{\citenamefont {Scully}\ and\ \citenamefont
  {Zubairy}(1997)}]{scully}%
  \BibitemOpen
  \bibfield  {author} {\bibinfo {author} {\bibfnamefont {M.~O.}\ \bibnamefont
  {Scully}}\ and\ \bibinfo {author} {\bibfnamefont {M.~S.}\ \bibnamefont
  {Zubairy}},\ }\href {\doibase 10.1017/CBO9780511813993} {\emph {\bibinfo
  {title} {Quantum Optics}}}\ (\bibinfo  {publisher} {Cambridge University
  Press},\ \bibinfo {year} {1997})\BibitemShut {NoStop}%
\bibitem [{\citenamefont {Breuer}\ and\ \citenamefont
  {Petruccione}(2006)}]{breuer}%
  \BibitemOpen
  \bibfield  {author} {\bibinfo {author} {\bibfnamefont {H.-P.}\ \bibnamefont
  {Breuer}}\ and\ \bibinfo {author} {\bibfnamefont {F.}~\bibnamefont
  {Petruccione}},\ }\href {\doibase 10.1093/acprof:oso/9780199213900.001.0001}
  {\emph {\bibinfo {title} {The Theory of Open Quantum Systems}}}\ (\bibinfo
  {year} {2006})\BibitemShut {NoStop}%
\bibitem [{\citenamefont {Zhao}\ \emph {et~al.}(2016)\citenamefont {Zhao},
  \citenamefont {De~Raedt}, \citenamefont {Miyashita}, \citenamefont {Jin},\
  and\ \citenamefont {Michielsen}}]{1}%
  \BibitemOpen
  \bibfield  {author} {\bibinfo {author} {\bibfnamefont {P.}~\bibnamefont
  {Zhao}}, \bibinfo {author} {\bibfnamefont {H.}~\bibnamefont {De~Raedt}},
  \bibinfo {author} {\bibfnamefont {S.}~\bibnamefont {Miyashita}}, \bibinfo
  {author} {\bibfnamefont {F.}~\bibnamefont {Jin}}, \ and\ \bibinfo {author}
  {\bibfnamefont {K.}~\bibnamefont {Michielsen}},\ }\bibfield  {title}
  {\enquote {\bibinfo {title} {Dynamics of open quantum spin systems: An
  assessment of the quantum master equation approach},}\ }\href {\doibase
  10.1103/PhysRevE.94.022126} {\bibfield  {journal} {\bibinfo  {journal} {Phys.
  Rev. E}\ }\textbf {\bibinfo {volume} {94}},\ \bibinfo {pages} {022126}
  (\bibinfo {year} {2016})}\BibitemShut {NoStop}%
\bibitem [{\citenamefont {Lages}\ \emph {et~al.}(2005)\citenamefont {Lages},
  \citenamefont {Dobrovitski}, \citenamefont {Katsnelson}, \citenamefont
  {De~Raedt},\ and\ \citenamefont {Harmon}}]{2}%
  \BibitemOpen
  \bibfield  {author} {\bibinfo {author} {\bibfnamefont {J.}~\bibnamefont
  {Lages}}, \bibinfo {author} {\bibfnamefont {V.~V.}\ \bibnamefont
  {Dobrovitski}}, \bibinfo {author} {\bibfnamefont {M.~I.}\ \bibnamefont
  {Katsnelson}}, \bibinfo {author} {\bibfnamefont {H.~A.}\ \bibnamefont
  {De~Raedt}}, \ and\ \bibinfo {author} {\bibfnamefont {B.~N.}\ \bibnamefont
  {Harmon}},\ }\bibfield  {title} {\enquote {\bibinfo {title} {Decoherence by a
  chaotic many-spin bath},}\ }\href {\doibase 10.1103/PhysRevE.72.026225}
  {\bibfield  {journal} {\bibinfo  {journal} {Phys. Rev. E}\ }\textbf {\bibinfo
  {volume} {72}},\ \bibinfo {pages} {026225} (\bibinfo {year}
  {2005})}\BibitemShut {NoStop}%
\bibitem [{\citenamefont {Esposito}\ and\ \citenamefont {Gaspard}(2003)}]{3}%
  \BibitemOpen
  \bibfield  {author} {\bibinfo {author} {\bibfnamefont {M.}~\bibnamefont
  {Esposito}}\ and\ \bibinfo {author} {\bibfnamefont {P.}~\bibnamefont
  {Gaspard}},\ }\bibfield  {title} {\enquote {\bibinfo {title} {Spin relaxation
  in a complex environment},}\ }\href {\doibase 10.1103/PhysRevE.68.066113}
  {\bibfield  {journal} {\bibinfo  {journal} {Phys. Rev. E}\ }\textbf {\bibinfo
  {volume} {68}},\ \bibinfo {pages} {066113} (\bibinfo {year}
  {2003})}\BibitemShut {NoStop}%
\bibitem [{\citenamefont {Weiss}(2012)}]{weiss}%
  \BibitemOpen
  \bibfield  {author} {\bibinfo {author} {\bibfnamefont {U.}~\bibnamefont
  {Weiss}},\ }\href {\doibase 10.1142/8334} {\emph {\bibinfo {title} {Quantum
  Dissipative Systems}}},\ \bibinfo {edition} {4th}\ ed.\ (\bibinfo
  {publisher} {World Scientific},\ \bibinfo {year} {2012})\BibitemShut
  {NoStop}%
\bibitem [{\citenamefont {Srednicki}(1999)}]{6}%
  \BibitemOpen
  \bibfield  {author} {\bibinfo {author} {\bibfnamefont {M.}~\bibnamefont
  {Srednicki}},\ }\bibfield  {title} {\enquote {\bibinfo {title} {The approach
  to thermal equilibrium in quantized chaotic systems},}\ }\href {\doibase
  10.1088/0305-4470/32/7/007} {\bibfield  {journal} {\bibinfo  {journal}
  {Journal of Physics A: Mathematical and General}\ }\textbf {\bibinfo {volume}
  {32}},\ \bibinfo {pages} {1163--1175} (\bibinfo {year} {1999})}\BibitemShut
  {NoStop}%
\bibitem [{\citenamefont {Balz}\ \emph {et~al.}(2018)\citenamefont {Balz},
  \citenamefont {Richter}, \citenamefont {Gemmer}, \citenamefont
  {Steinigeweg},\ and\ \citenamefont {Reimann}}]{balz}%
  \BibitemOpen
  \bibfield  {author} {\bibinfo {author} {\bibfnamefont {B.~N.}\ \bibnamefont
  {Balz}}, \bibinfo {author} {\bibfnamefont {J.}~\bibnamefont {Richter}},
  \bibinfo {author} {\bibfnamefont {J.}~\bibnamefont {Gemmer}}, \bibinfo
  {author} {\bibfnamefont {R.}~\bibnamefont {Steinigeweg}}, \ and\ \bibinfo
  {author} {\bibfnamefont {P.}~\bibnamefont {Reimann}},\ }\enquote {\bibinfo
  {title} {Dynamical typicality for initial states with a preset measurement
  statistics of several commuting observables},}\ in\ \href {\doibase
  10.1007/978-3-319-99046-0_17} {\emph {\bibinfo {booktitle} {Thermodynamics in
  the Quantum Regime: Fundamental Aspects and New Directions}}}\ (\bibinfo
  {publisher} {Springer International Publishing},\ \bibinfo {address} {Cham},\
  \bibinfo {year} {2018})\ pp.\ \bibinfo {pages} {413--433}\BibitemShut
  {NoStop}%
\bibitem [{\citenamefont {Reimann}\ and\ \citenamefont
  {Gemmer}(2019)}]{physica}%
  \BibitemOpen
  \bibfield  {author} {\bibinfo {author} {\bibfnamefont {P.}~\bibnamefont
  {Reimann}}\ and\ \bibinfo {author} {\bibfnamefont {J.}~\bibnamefont
  {Gemmer}},\ }\bibfield  {title} {\enquote {\bibinfo {title} {Why are
  macroscopic experiments reproducible? {I}mitating the behavior of an ensemble
  by single pure states},}\ }\href {\doibase 10.1016/j.physa.2019.121840}
  {\bibfield  {journal} {\bibinfo  {journal} {Physica}\ }\textbf {\bibinfo
  {volume} {A}},\ \bibinfo {pages} {121840} (\bibinfo {year}
  {2019})}\BibitemShut {NoStop}%
\bibitem [{\citenamefont {Reimann}(2018)}]{limits}%
  \BibitemOpen
  \bibfield  {author} {\bibinfo {author} {\bibfnamefont {P.}~\bibnamefont
  {Reimann}},\ }\bibfield  {title} {\enquote {\bibinfo {title} {Dynamical
  typicality of isolated many-body quantum systems},}\ }\href {\doibase
  10.1103/PhysRevE.97.062129} {\bibfield  {journal} {\bibinfo  {journal} {Phys.
  Rev. E}\ }\textbf {\bibinfo {volume} {97}},\ \bibinfo {pages} {062129}
  (\bibinfo {year} {2018})}\BibitemShut {NoStop}%
\bibitem [{\citenamefont {Tal-Ezer}\ and\ \citenamefont
  {Kosloff}(1984)}]{chev1}%
  \BibitemOpen
  \bibfield  {author} {\bibinfo {author} {\bibfnamefont {H.}~\bibnamefont
  {Tal-Ezer}}\ and\ \bibinfo {author} {\bibfnamefont {R.}~\bibnamefont
  {Kosloff}},\ }\bibfield  {title} {\enquote {\bibinfo {title} {An accurate and
  efficient scheme for propagating the time dependent {S}chr\"odinger
  equation},}\ }\href {\doibase 10.1063/1.448136} {\bibfield  {journal}
  {\bibinfo  {journal} {The Journal of Chemical Physics}\ }\textbf {\bibinfo
  {volume} {81}},\ \bibinfo {pages} {3967--3971} (\bibinfo {year}
  {1984})}\BibitemShut {NoStop}%
\bibitem [{\citenamefont {Raedt}\ and\ \citenamefont
  {Michielsen}(2004)}]{chev2}%
  \BibitemOpen
  \bibfield  {author} {\bibinfo {author} {\bibfnamefont {H.~De}\ \bibnamefont
  {Raedt}}\ and\ \bibinfo {author} {\bibfnamefont {K.}~\bibnamefont
  {Michielsen}},\ }\href {https://arxiv.org/abs/quant-ph/0406210v2} {\emph
  {\bibinfo {title} {Computational Methods for Simulating Quantum Computers}}}\
  (\bibinfo  {publisher} {American Scientific Publisher},\ \bibinfo {year}
  {2004})\BibitemShut {NoStop}%
\end{thebibliography}%
\appendix
\section*{Appendix}
\addcontentsline{toc}{section}{Appendices}
\renewcommand{\theequation}{\thesubsection.\arabic{equation}}
\setcounter{equation}{0}
\renewcommand{\thesubsection}{\Alph{subsection}}
Details of our numerical implementation will be discussed in this appendix. We make use of the concept of typicality (cf. App. \ref{typ}) and use a time evolution algorithm (real and imaginary) based on Chebyshev polynomials (cf. App. \ref{chev}). Since the full Hamiltonian conserves magnetization, we perform the procedure outlined below in each magnetization subspace. The dynamics in the full Hilbert space are obtained by piecing together the contributions of each magnetization subspace weighted with their respective binomial weight. Note that the main hindrance to our calculations is not the exponentially large Hilbert space dimension, but rather the extremely long times that have to be reached in real time. In App. \ref{gene} a result for randomized bath couplings is shown.
\subsection{Typicality}
\label{typ}
As mentioned in Sect. \ref{spinexp}, the initial state is a product state of a microcanonical bath state and a projected spin-up system state. Since the numerical integration of the von-Neumann equation can be cumbersome, we make use of the concept of typicality, which states that a single ``typical'' pure state can have the same thermodynamic properties as the full statistical ensemble \cite{balz}. Not only is it more memory efficient to work with pure states, the availability of efficient time evolution algorithms for pure states, e.g. Runge-Kutta or Chebyshev polynomials, constitutes a major advantage. To find such a typical state, a pure state $|\phi\rangle$ is drawn at random from the Hilbert space according to the unitary invariant Haar measure.
\begin{equation}
|\phi\rangle=\sum_i c_i|i\rangle
\end{equation}
Real and imaginary part of the complex coefficients $c_i$ are drawn from a Gaussian distribution with mean zero and unit variance and the set $\{|i\rangle\}$ is an arbitrary basis of the Hilbert space, e.g. the Ising basis.
Consider the new normalized state
\begin{equation}
\label{typp}
|\psi\rangle=\dfrac{\sqrt{\rho}\,|\phi\rangle}{\sqrt{\langle\phi|\rho|\phi\rangle}}\,.
\end{equation}
It can be shown \cite{physica} that for the overwhelmingly majority of random states $|\phi\rangle$, the pure state $|\psi\rangle$ exhibits effectively the the same thermodynamic behavior as the mixed state $\rho$, i.e.
\begin{equation}
\langle A(t)\rangle = \text{Tr}\{\rho A(t)\} = \langle \psi | A(t) | \psi \rangle + \epsilon\,.
\end{equation}
%essentially replacing the trace $\text{Tr}\{\bullet\}$ by a scalar product $\langle\psi|\bullet|\psi\rangle$. 
Importantly, the induced error $\epsilon =\epsilon(|\psi\rangle)$ has mean zero, i.e. $\overline{\epsilon} =0$, and a standard deviation that scales inversely proportional to the square root of the effective Hilbert space dimension, i.e. $\sigma(\epsilon) \propto 1/\sqrt{d_{\text{eff}}}$ \cite{limits}. The effective dimension $d_{\text{eff}} = 1/\text{Tr}\{\rho^2\}$ is a measure of how many pure states contribute to the mixture $\rho$.\newline
In the paper at hand the initial state $\rho$ (cf. Eq. (\ref{ini})) is a projection operator. Therefore, it is permissible to drop the square root in the numerator in Eq. (\ref{typp}). Now the projectors $\pi_{\uparrow} \otimes 1$ and $1 \otimes \pi_{E,\delta}$ need to be applied to the state $|\phi\rangle$, which is an element of the product Hilbert space. As we are working in the Ising basis, the action of $\pi_{\uparrow} \otimes 1$ on $|\phi\rangle$ is easily implemented by setting corresponding components of the state vector to zero. Since it is unfeasible to diagonalize the full many-body Hamiltonian, we replace the bath projector by a Gaussian filter which suppresses contributions of energy eigenstates far away from the desired energy $E$, resulting in a narrowly populated energy window of width (variance) $\delta$.
\begin{equation}
\pi_{E,\delta} \approx \text{exp}\bigg{(}{\dfrac{-(H_{\text{bath}}-E)^2}{2 \vphantom{\hat{\delta}}\delta}}\bigg{)}
\end{equation}
The Gaussian filter is applied with a Chebyshev algorithm (cf. App. \ref{chev}). Lastly, the resulting wave function is normalized to obtain the state $|\psi\rangle$ as in Eq. (\ref{typp}). In this scenario the effective dimension $d_{\text{eff}}$ is essentially the number of states in the energy window. Increasing the size of the system, while keeping $\delta$ fixed, results in an exponentially growing effective dimension $d_{\text{eff}}$ and therefore in a negligible typicality error for moderately sized systems.

\subsection{Chebyshev polynomials}
\label{chev}
\setcounter{equation}{0}
A Chebyshev type algorithm is employed in order to evolve a pure state in real and imaginary time \cite{chev1,chev2}.
Say it is desirable to approximate a scalar function $f(x)$ in the interval $[-1,1]$ by a polynomial expansion, i.e.
\begin{equation}
f(x) \approx \sum_n c_n P_n(x)\,,
\end{equation}
with coefficients $c_n$ and polynomials $P_n$ of order $n$. The unique set of polynomials that minimizes the maximum error in this interval is called Chebyshev polynomials of the first kind. They are denoted by $T_n(x)$ and can be written down recursively as
\begin{equation}
\label{iter}
T_{n}(x)= 2x T_{n-1}(x) - T_{n-2}(x)
\end{equation}
with $T_0(x)=1$ and $T_1(x)=x$. They are orthogonal with respect to the weighted scalar product
\begin{equation}
\langle T_n|T_m \rangle = \int_{-1}^{1}\dfrac{T_n(x)T_m(x)}{\pi \sqrt{1-x^{2}}}\,\text{d}x = \delta_{nm} C_n
\end{equation}
with $C_0=1$ and $C_{n>0}=1/2$. In order to approximate the time evolution operator, the bandwidth of the Hamiltonian has to be rescaled accordingly. Defining $a=(E_{\text{max}}-E_{\text{min}})/2$ and $b=(E_{\text{max}}+E_{\text{min}})/2$, where $E_{\text{max}}$ ($E_{\text{min}}$) is the maximal (minimal) energy eigenvalue, the rescaled Hamiltonian is obtained by $\tilde{H} = (H-b)/a$. In practice a small safety parameter is chosen that ensures that the rescaled spectrum lies well within $[-1,1]$. Now we get
\begin{equation}
\label{cheby}
e^{-\mathrm{i} H \Delta t}=e^{-\mathrm{i} b \Delta t} \bigg{[} c_0(a\Delta t)+2\sum_{n \geq1} c_n(a\Delta t) T_n(\tilde{H})\bigg{]}
\end{equation}
with complex coefficients 
\begin{equation}
c_n(a\Delta t)=\int_{-1}^{1}\dfrac{T_n(x)e^{-\mathrm{i}xa\Delta t}}{\pi \sqrt{1-x^2}}\text{d}x=(-\mathrm{i})^n J_n(a\Delta t),
\end{equation}
 where $J_n$ denotes the $n$-th order Bessel function of the first kind. Since the coefficients only depend on the time step, but not on time itself, they only have to calculated once. Applying Eq. (\ref{cheby}) to a state $|\psi(t)\rangle$ boils down to calculating $T_n(\tilde{H})|\psi(t)\rangle$ for various $n$, which can be done iteratively using Eq. (\ref{iter}). Terminating the sum in Eq. (\ref{cheby}) at an upper bound $M$ gives the $M$-th order Chebyshev approximation of the time evolution operator. The required order for convergence depends on the particular problem. For our biggest system with $N=25$ spins we had to go up to order $40$.

 \subsection{Validity of the numercial results for a lager class of systems}
\label{gene}
While the investigated model class may seem peculiar, it has just been chosen as one generic representative of the whole of condensed matter type systems. 
To exclude that overall results are just due to  any unintentional, subtle  conserved quantities, etc., we redid parts of our numerical analysis with randomized 
bath couplings,
drawn from a Gaussian distribution with mean zero and standard deviation $0.2$. 
One result is displayed in Fig. \ref{rand} for $N=25$ and $\lambda = 0.2$. 
One readily verifies that the curves coincide nicely. This hints at the generic nature of our model class.
\begin{figure}[H]
	\centering
  \includegraphics[width=0.48\textwidth]{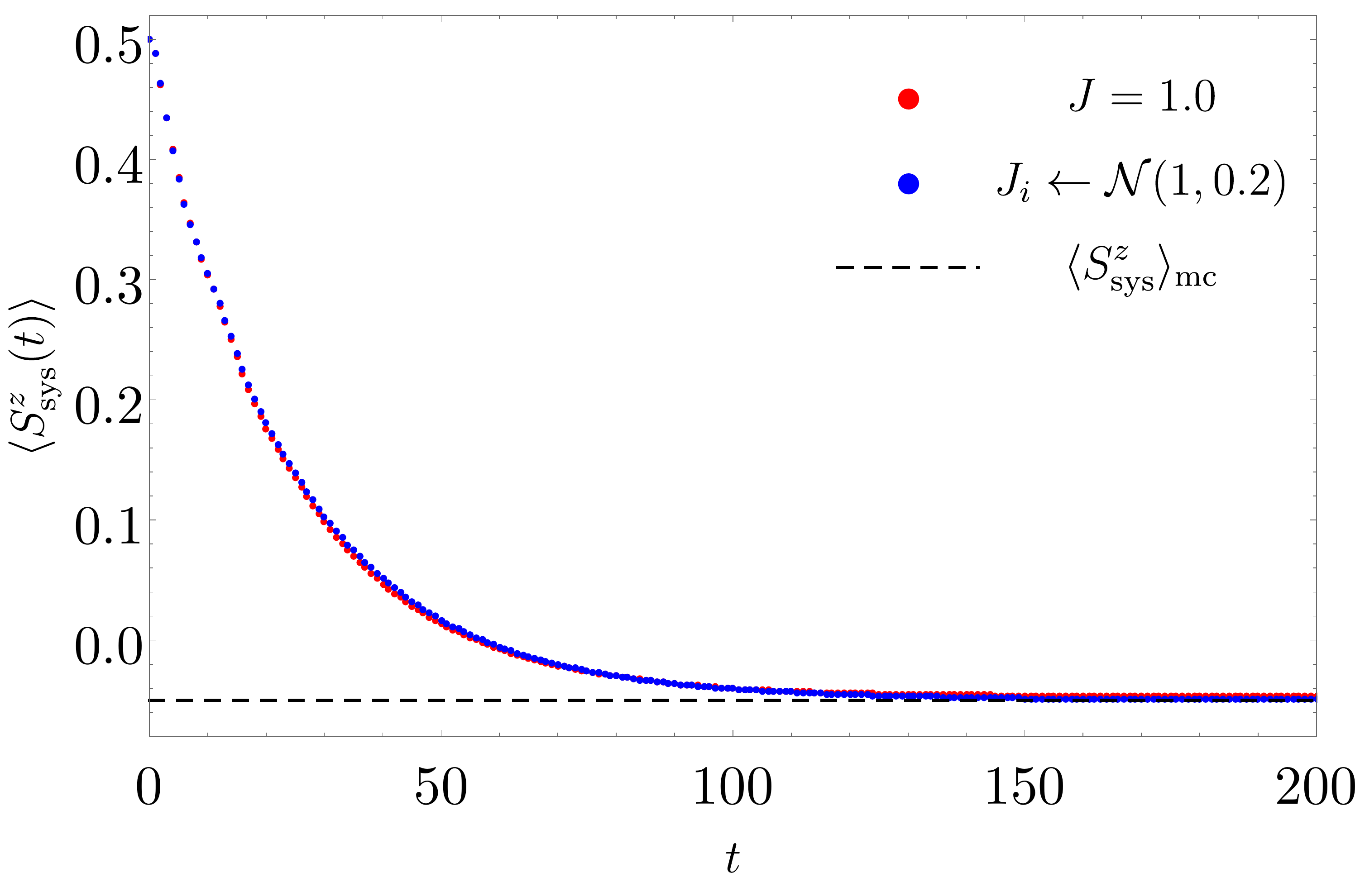}
	\caption{Comparison between setups with couplings set to unity and randomly drawn couplings. There is no apparent difference.}
	\label{rand}
\end{figure}
\newpage
\end{document}